\documentclass[a4paper,11pt]{article}

\pdfoutput=1
\usepackage{jcappub}

\usepackage{graphicx}
\usepackage{longtable}
\usepackage{float}
\usepackage{dcolumn}
\usepackage{bm}
\usepackage{appendix}
\usepackage{multirow}
\usepackage{color}
\usepackage[utf8]{inputenc}
\usepackage{footmisc}
\usepackage{hyperref}

\usepackage{txfonts}
\usepackage{xcolor}
\usepackage{graphicx}
\usepackage{bm}
\usepackage{ulem}
\usepackage{multirow, array, float, tabularx}
\usepackage{amsmath}
\hypersetup{
	citecolor = red,
	linkcolor = blue,
	urlcolor = magenta
}

\newcommand{\be}{\begin{equation}}
	\newcommand{\ee}{\end{equation}}
\newcommand{\bea}{\begin{eqnarray}}
	\newcommand{\eea}{\end{eqnarray}}
\newcommand{\bes}{\begin{subequations}}
	\newcommand{\ees}{\end{subequations}}

\newcommand{\bc}{\begin{center}}
	\newcommand{\ec}{\end{center}}

\usepackage{wasysym}

\begin{document}

	\title{Constraints on the non-minimally coupled Witten-O'Raifeartaigh inflation}
	
	\author[a]{F. B. M. dos Santos}\emailAdd{felipebrunomedeiros@gmail.com}
    \author[a,b]{R. Silva}\emailAdd{raimundosilva@fisica.ufrn.br}
    \author[c]{J. S. Alcaniz}\emailAdd{alcaniz@on.br}

    \affiliation[a]{Universidade Federal do Rio Grande do Norte, Departamento de F\'{\i}sica, Natal - RN, 59072-970, Brasil}

    \affiliation[b]{Departamento de F\'{\i}sica, Universidade do Estado do Rio Grande do Norte, Mossor\'o, 59610-210, Brasil}

	\affiliation[c]{Observatório Nacional, Rua General José Cristino 20921-400, Rio de Janeiro-RJ, Brasil}
	
	\abstract{We investigate the impact of a non-minimal coupling of the scalar field with gravity in inflationary models, where a small coupling is allowed. As a concrete example, we consider the Witten-O'Raifeartaigh model, where, in line with other models, the presence of a coupling strength $\xi$ can recover concordance of the inflationary parameters with cosmic microwave background (CMB) constraints, provided by the Planck collaboration. We go beyond the slow-roll regime and investigate the impact in the description of CMB anisotropies by performing a statistical analysis of the model with the most recent Planck + Baryon Acoustic Oscillations (BAO) data to seek for any indication of a non-zero coupling by data within the model. We find that not only the presence of a non-minimal coupling is seen, but the model has a slight statistical preference when compared with the standard $\Lambda$CDM one. We also discuss the results on the minimally-coupled model, which in general, favours the simple setting where the associated mass scale is equal to the reduced Planck mass $M_p$ while being, in general, disfavored concerning the standard model.}

	\maketitle

	\section{Introduction}\label{sec1}

	The existence of a period of acceleration in the very early universe, known as cosmic inflation \cite{Guth1981,Starobinsky:1980te,Linde:1981mu,Linde:1983gd} is seriously considered a realistic description of the primordial era, as it solves associated problems with the standard Big Bang picture, at the same time that it provides a mechanism for the growth of fluctuations in the subsequent evolution of the universe. From the observational standpoint, observations of the cosmic microwave background (CMB) \cite{Mather:1993ij,WMAP:2003syu,WMAP:2006bqn} seem to favour this epoch in cosmic history; in particular, the most recent Planck collaboration results \cite{Aghanim:2018eyx} give strong constraints on cosmological parameters in a manner that it is possible to test the viability of different scenarios. As for what causes the universe to expand in an accelerated state, one of the most straightforward explanations is the presence of a scalar field with a potential function that dictates its evolution. This potential must generally have a region with a low enough curvature so that the field evolves slowly and there is little change in the potential energy during inflation. It characterizes the slow-roll regime, and after its breakdown, a period of reheating \cite{Abbott:1982hn,Albrecht:1982mp,Kofman:1994rk} must take place, where the leftover energy of the field is converted to other particles, allowing the subsequent eras of radiation and matter to happen.  
	
	Although there is a wide variety of possible scenarios \cite{Martin:2013tda,Martin:2013nzq}, one interesting aspect is that the simplest models that can be constructed, such as the ones characterized by a quartic or an exponential potential, fail to meet the restrictions imposed by CMB data, in the $n_s-r$ plane. In this manner, one can search for alternatives to bring such models back to viability. One way to achieve this is by considering that the scalar field is non-minimally coupled to gravity in the form $\xi\phi^2R/2$ \cite{Lucchin:1985ip,Futamase:1987ua,Fakir1990,Komatsu1998,Komatsu1999,Linde:2011nh}, with $\xi$ determining how strong this coupling is. Over the years, there have been extensive studies on the viability of such models, in which it can be found that concordance with data can be restored not only for the simplest models but for more complex ones as well \cite{Lucchin:1985ip,Futamase:1987ua,Komatsu1999,Komatsu1998,Fakir1990,Hertzberg:2010dc,Linde:2011nh,Okada:2014lxa,Kaiser:2015usz,Bezrukov2008,Tenkanen:2017jih,Campista:2017ovq,Ferreira:2018nav,Bostan:2018evz,Reyimuaji:2020goi,Rodrigues:2021txa,Rodrigues:2020dod,dosSantos:2021vis}.
	
	In this particular work, we seek to establish constraints on non-minimally coupled models by extending the analysis done in \cite{Santos:2022aeb}, in which the Witten-O'Raifeartaigh (WR) model of inflation \cite{Albrecht:1983ib,Martin:2013tda} was investigated in the context of a non-minimally coupled scalar field. In the original picture, the model emerges from supergravity \cite{Albrecht:1983ib,Witten:1981kv,Witten:1981nf}, in an attempt to solve the hierarchy problem, such that the mass scale associated with the model is estimated to be of the order of Planck mass. On the other hand, it is also possible within the supergravity context to arrive at the same model through a different construction, where the mass scale associated is not theoretically constrained a priori \cite{Artymowski:2019jlh}. A discussion of the model in the context of a non-minimally coupled field to gravity was done \cite{Santos:2022aeb}, whereby following previous works \cite{Martin:2013nzq}, a wide range of the mass $M$ was considered. The main result is that in order for the model to have a good concordance with the recent CMB constraints on the $n_s-r$ parameters, the strength of the non-minimal coupling must be of the order of $\xi\sim 10^{-3}$. This implies that one could, in principle, impose a precise estimate on this parameter through a complete analysis with the full CMB likelihood. We note that this result contrasts with other realizations of non-minimally coupled inflation models since a strong coupling of the field seems to emerge depending on the model considered. In Higgs inflation \cite{Bezrukov2008}, a coupling of order $\xi\sim 10^4$ makes the model compatible with data; also, when radiative corrections to the scalar potential are considered \cite{Rodrigues:2020dod,Rodrigues:2021txa}, a robust coupling of the field with gravity is necessary. This feature motivates further numerical analyses, in which one can quantify how relevant such an extension of gravity is in the early universe. 
	
	We use Planck 2018+BAO data to establish constraints on the non-minimally coupled WR model, with two main objectives: One is to verify a possible presence of a non-minimal coupling of the field with gravity, checking if the strength of the coupling is compatible with the restrictions on the inflationary parameters. The other is to check for any preference of this extended model concerning the standard $\Lambda$CDM one. Although it is expected that a better fit to data can be achieved if extra parameters are considered, statistical criteria tend to penalize complex models so that various possible scenarios can be discarded as viable ones. We also remember that the model has an associated mass scale $M$, so it is instructive to see the impact of a varying $M$.
	
	This work is organized in the following manner: Section \ref{sec2} briefly reviews the formalism for a non-minimally coupled field and the main results for the non-minimally coupled WR model in the Einstein frame. In Section \ref{sec3}, we describe the methods and data used for the statistical analysis, while in Section \ref{sec4}, we discuss the results obtained. Finally, Section \ref{sec5} presents our considerations.
	
	\section{Non-minimally coupled slow-roll inflation}\label{sec2}
	
	This section reviews the slow-roll analysis for a non-minimally coupled scalar field and its application to the WR model.
	
	\subsection{Formalism}
	
	When considering a non-minimally coupled scalar field with gravity, one can start with the following action
	\begin{gather}
		S=\int d^4x\sqrt{-g}\left[\frac{\Omega^2(\phi)}{16\pi G}R - \frac{1}{2}g^{\mu\nu}\partial_\mu\phi\partial_\nu\phi - V(\phi)\right],
		\label{eq:2.1}
	\end{gather}
	with $\Omega^2(\phi)$ being the function that characterizes the coupling of the field with gravity, while $\frac{1}{\sqrt{8\pi G}}=M_p$ is the reduced Planck mass. By choosing $\Omega^2(\phi)=1+8\pi G\xi\phi^2$, we find that $\xi$ is a dimensionless constant corresponding to the strength of the coupling, with the $\xi=0$ case reproducing the dynamics of a minimally coupled field. To simplify the following analysis, we choose to work in the Einstein frame, in which the action (\ref{eq:2.1}) can be written as one correspondent to a minimally coupled scalar field, as
	\begin{gather}
		S = \int d^4x\sqrt{-g}\left[\frac{\tilde R}{16\pi G}-\frac{1}{2}\tilde{g}^{\mu\nu}\partial_\mu\chi\partial_\nu\chi-\tilde{V}(\chi)\right],
		\label{eq:2.2}
	\end{gather}
	resulting from the transformation $\tilde{g}_{\mu\nu}=\Omega^2(\phi)g_{\mu\nu}$, and by a redefinition of the field as
	\begin{gather}
		\frac{d\chi}{d\phi}=\frac{\sqrt{1+\xi\phi^2(1+6\xi)/M_p^2}}{\left(1+\xi\phi^2/M_p^2\right)}.
		\label{eq:2.3}
	\end{gather}
	Also, the Einstein frame potential is expressed as
	\begin{gather}
		\tilde{V}(\chi)\equiv\frac{V(\phi)}{\left(1+\xi\frac{\phi^2}{M_p^2}\right)^2}.
		\label{eq:2.4}
	\end{gather}

	\begin{figure*}
		\centering
		\includegraphics[width=0.7\columnwidth]{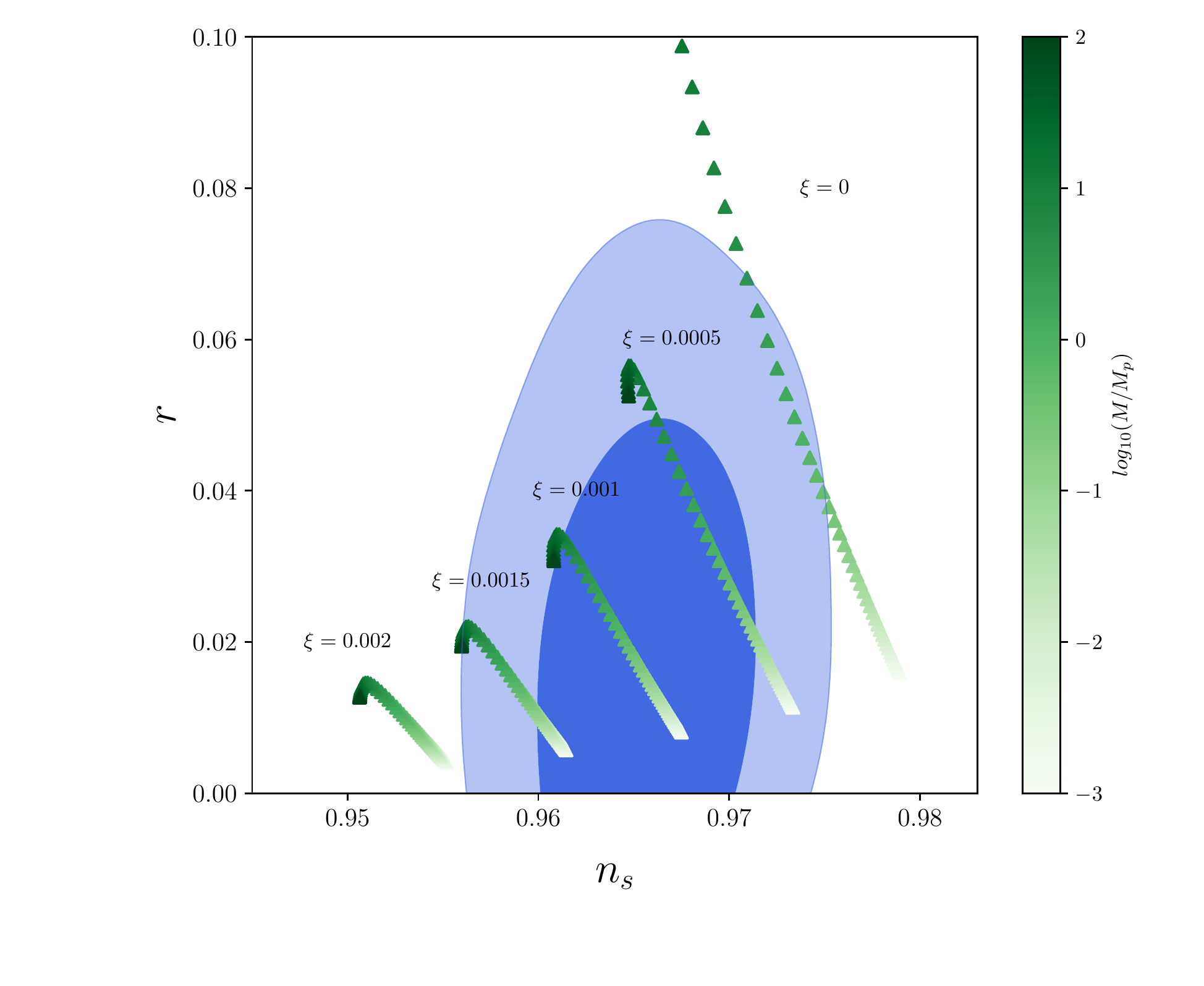}
		\caption{The $n_s-r$ predictions for the non-minimally coupled WR model, with the most recent constraints from Planck TT + lowP + BAO + BK15 \cite{Aghanim:2018eyx}. The color bar shows the values of $\log_{10}M/M_p$, while we choose fixed values of $\xi$, indicated in the picture. We also take $N_\star=55$.}
		\label{fig:1}
	\end{figure*}
	
	In order to extract predictions from a given model, we first remember that the amplitudes of the scalar and tensor power spectra are respectively given as
	\begin{gather}
		\mathcal{P}_\mathcal{R}=\frac{V}{24\pi^2M_p^4\epsilon}\Bigg|_{k=k_{\star}}, \quad \mathcal{P}_t=\frac{H^2}{2\pi^2M_p^2}\Bigg|_{k=k_{\star}},
		\label{eq:2.5}
	\end{gather}
	in a manner that the variation of $\mathcal{P}_\mathcal{R}$ with scale and the ratio $\mathcal{P}_t/\mathcal{P}_\mathcal{R}$ give the well-known spectral index and tensor-to-scalar ratio, respectively, both at $k=k_\star$, the scale at which the CMB scale leaves the horizon: 
	\begin{gather}
		n_{s}=1-6\epsilon_\star + 2\eta_\star, \quad r=16\epsilon_\star,
		\label{eq:2.6}
	\end{gather}
	
	We remember the observational values given by recent CMB-dedicated experiments. The primordial spectrum amplitude $P_{R}$ at the pivot scale is given by the \textit{Planck} collaboration as $\log(10^{10}\mathcal{P}_\mathcal{R})=3.044\pm0.014$. The tensor-to-scalar ratio has today an upper limit of $r<0.056$, while the spectral index is constrained as $n_s=0.9649 \pm 0.0042$, at $68\%$ confidence level \cite{Planck:2018jri}.
		
	\subsection{Non-minimally coupled WR model}

	\begin{figure*}
		\centering
		\includegraphics[width=0.49\columnwidth]{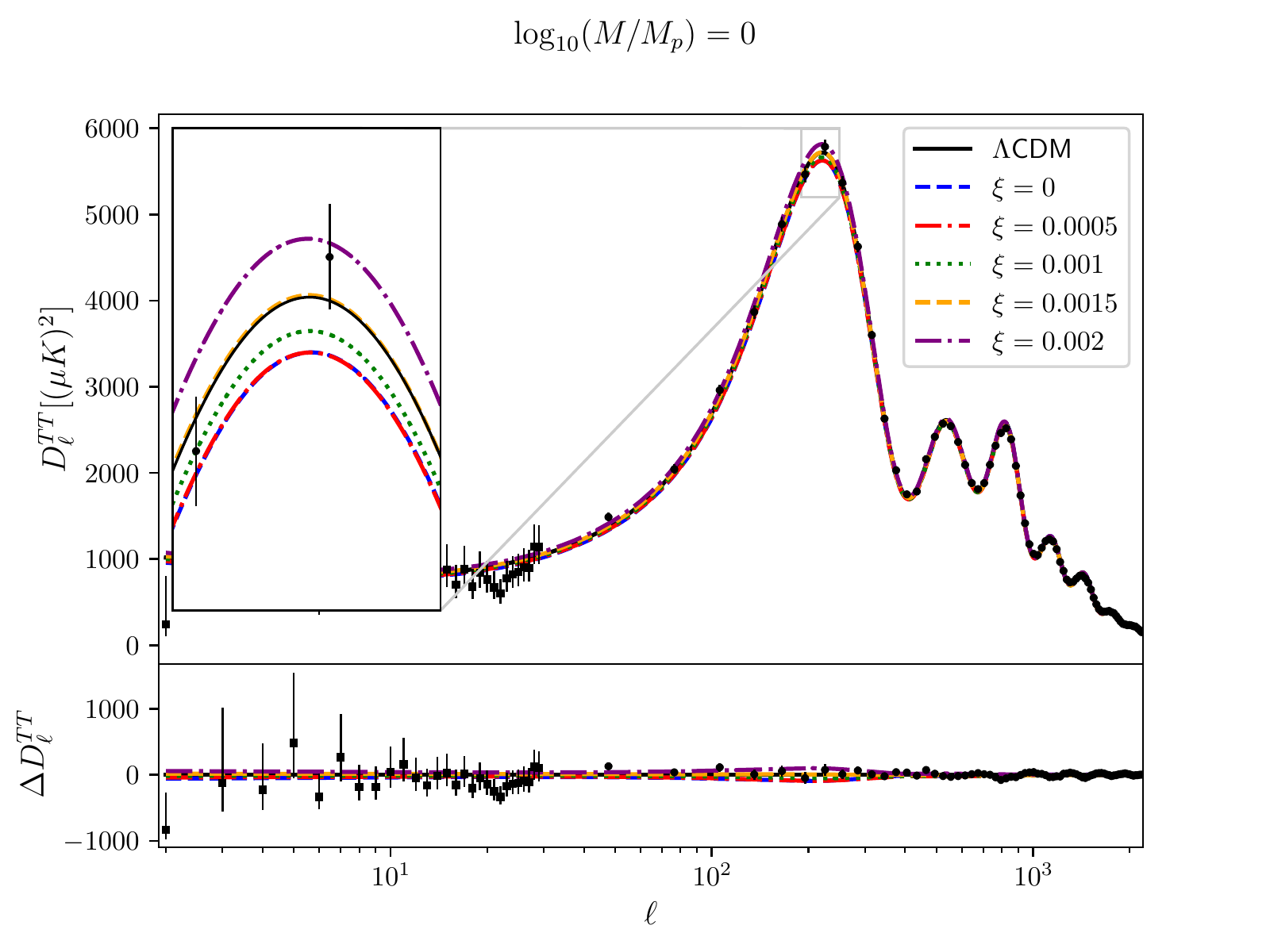}
		\includegraphics[width=0.49\columnwidth]{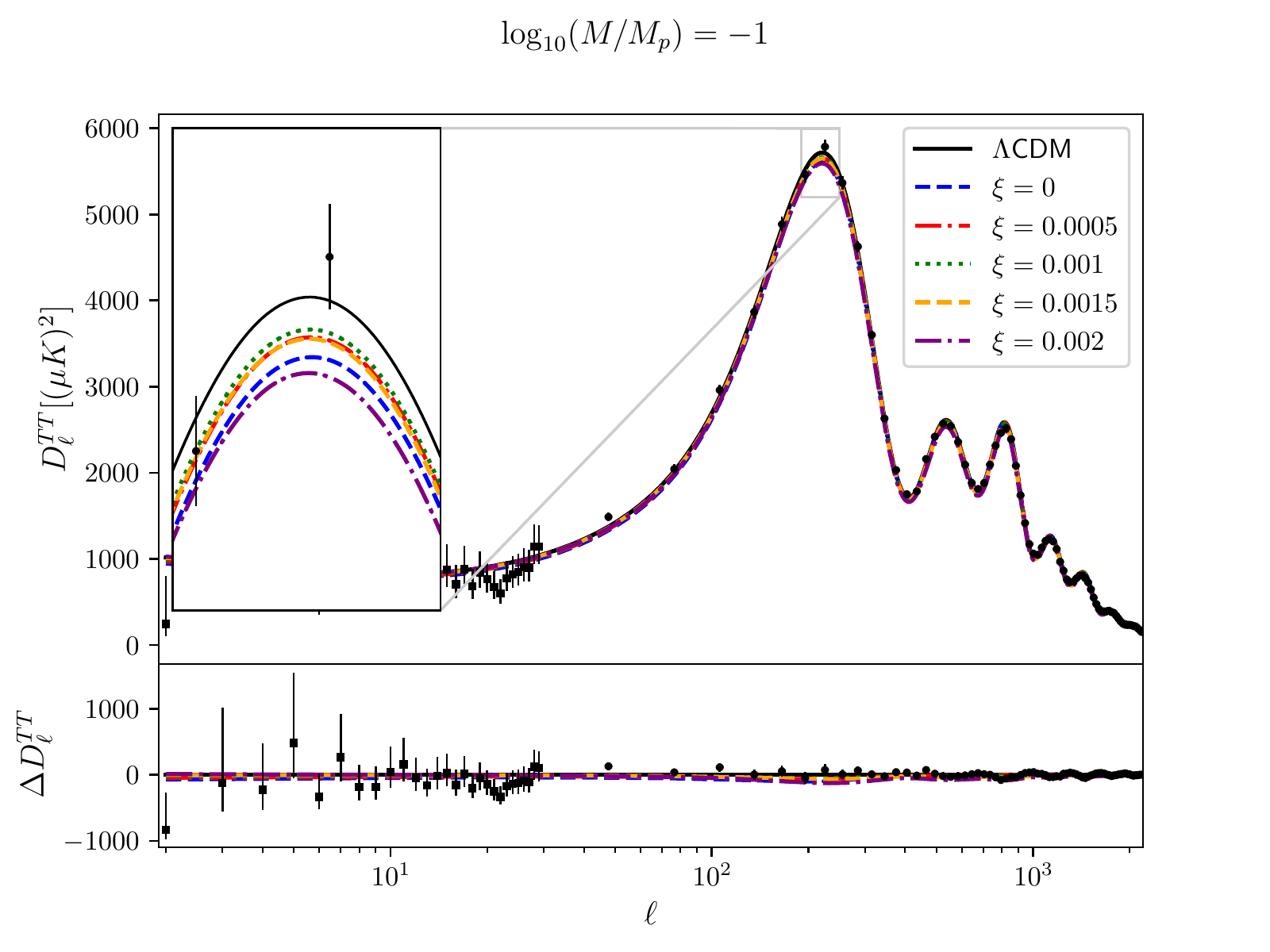}
		\includegraphics[width=0.49\columnwidth]{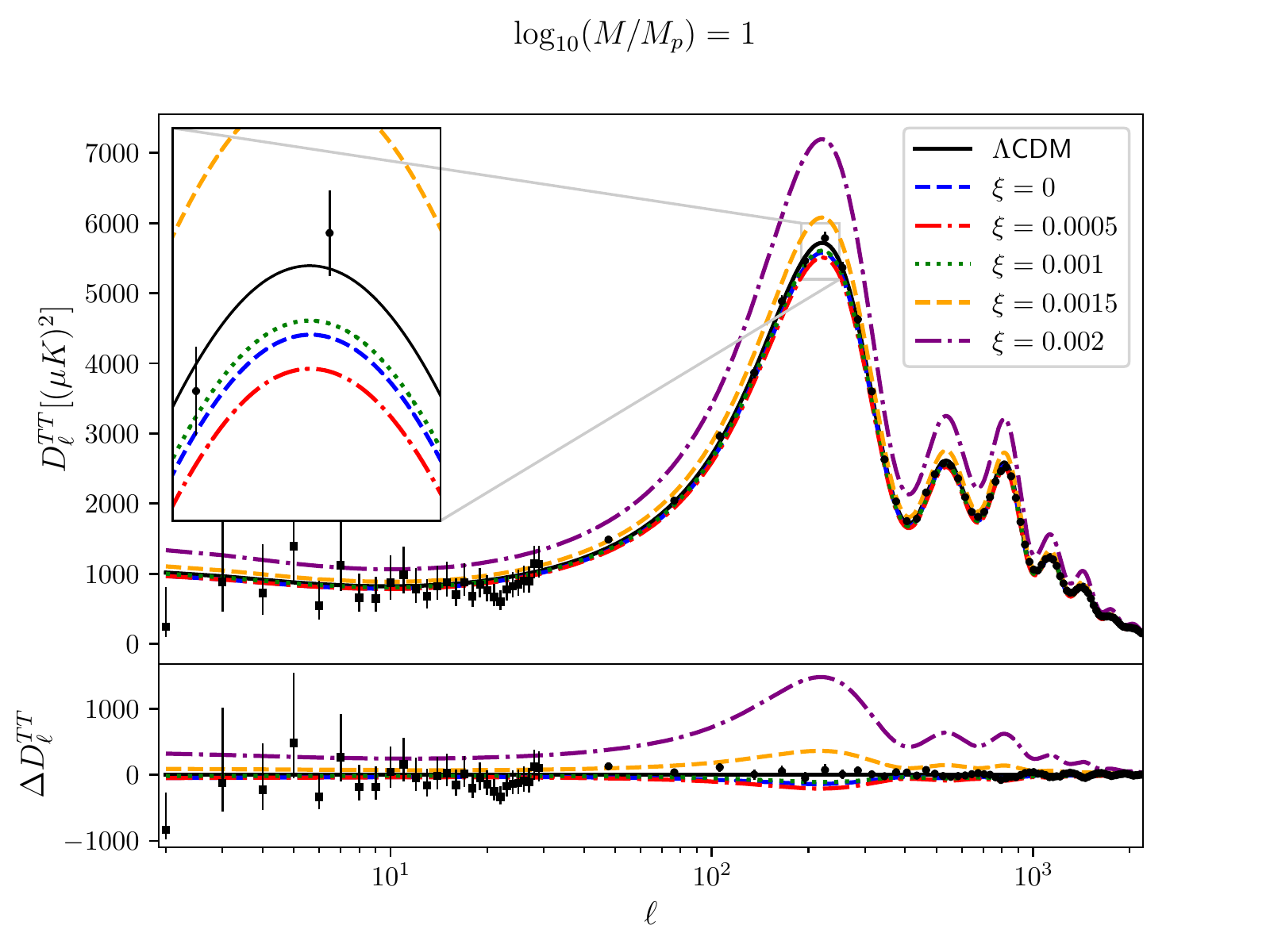}
		\caption{The CMB temperature power spectra for the non-minimally coupled WR model, together with Planck data \cite{Aghanim:2018eyx}. We have considered fixed values of $M$, being $\log_{10}(M/M_p)=0$ (upper left panel), $\log_{10}(M/M_p)=-1$ (upper right panel) and $\log_{10}(M/M_p)=1$ (lower panel), while the parameter $\xi$ spans the range $\xi=[0 : 0.002]$.}
		\label{fig:2}
	\end{figure*} 
	
The potential characterizing the WR model is given by
	\begin{gather}
		V(\phi) = V_0\log^2\left(\frac{\phi}{M}\right),
        \label{eq:2.7}
	\end{gather}
	where $V_0$ represents the amplitude of the potential, while $M$ is a mass scale that controls its minimum value. The model was investigated in detail for $\xi\geq 0$ in \cite{Santos:2022aeb}, where the Einstein frame potential is given by Eq. (\ref{eq:2.4}) and we revise the main results on the predictions regarding the $n_s-r$ plane. It is shown in fig. \ref{fig:1}. We have chosen $\xi=0,0.0	005,0.001,0.0015,0.002$, varying $M$ in a large interval, as shown in the color bar. In general, an increasing $M$ leads to a higher tensor-to-scalar ratio, a feature that is easy to see in the minimally coupled case; for $\xi\neq 0$, however, this behaviour changes, and $r$ will start decreasing for an increasing $M$ at some point, estimated as being approximately $M\sim 10 M_p$. More importantly, it is possible to note that a small $\xi$ is sufficient to make the predictions fit into the Planck restrictions, even if we consider a wide range of $M$, and that higher $\xi$ leads to smaller $n_s$, which helps us in determining the viable range of the non-minimal coupling.
	
	The $\xi$ parameter also affects the CMB temperature power spectrum. In fig. \ref{fig:2}, we see the non-minimally coupled WR model (NMWR) predictions. First, we have fixed $M=M_p$ (upper left panel) and varied only $\xi$; we immediately see the effect of an increasing $\xi$,  which is to increase the amplitude of the spectrum. For a smaller $M=0.1M_p$ (upper right panel), we notice a sudden decrease in the amplitude for all $\xi$ concerning the $\Lambda$CDM prediction (black line). The increase of $M$, as given in the lower panel ($M=10M_p$), results in a better fit with data when smaller $\xi$ are considered; in this manner, the correlation between the model parameters and the increase/decrease of the spectra is non-trivial, implying that only a complete analysis will tell us the real impact of a non-minimal coupling to the model. From this initial analysis, however, we then note two main features: First, a small coupling of order $\xi\sim 10^{-3}$ has a very significant effect on the amplitude of the spectra, which might indicate that it is possible to constrain this parameter with reasonable precision. Second, the mass scale $M$ also plays an important role, as it helps to suppress or increase the effect of $\xi$ in the predictions. This is essential, as the role of $M$ here in the non-minimally coupled case is different than in the $\xi=0$ one, as the curves are generally similar to each other for the interval of $M$ considered (blue curves).

 \begin{table}[t]
		\begin{center}
			\begin{tabular}{|c | c|} 
				\hline
				Parameter & Priors  \\ [0.5ex] 
				\hline
				\hline
				$\Omega_bh^2$ &[0.005 : 0.1]  \\ 
				
				$\Omega_ch^2$ & [0.001 : 0.99] \\
				
				$100\theta$ & [0.5 : 10] \\
				
				$\tau$ & [0.01 : 0.8]  \\
				
				$\xi$ & [0 : 0.003] \\
				
				$\log_{10}(M/M_p)$ & [-3 : 2] \\
				\hline
			\end{tabular}
			\caption{Priors used in the statistical analysis.}
			\label{tab:1}
		\end{center}
	\end{table}

	\section{Methodology and Analysis}\label{sec3}
	
	In the previous section, we have seen how greatly a non-minimal coupling to gravity impacts the predictions of the WR model, essentially bringing the parameters $n_s$ and $r$ into the Planck 2018+BAO constraints for a large interval of $M$. This section describes the method we use to investigate the model further. Thus, we perform a numerical analysis to compare the theoretical predictions of the model with the CMB data reported in the Planck collaboration. To this end, we use a modification of the Code for Anisotropies in the Microwave Background (CAMB) \cite{Lewis:1999bs,Lewis:2002ah}, included in the \textsc{ModeCode} \cite{Mortonson:2010er} extension, to solve the Boltzmann equations for a particular scalar field model, and produce their respective power spectra. It is necessary, since CAMB assumes a simpler power-law form for the scalar power spectrum, $\mathcal{P}_\mathcal{R}=A_s\left(k/k_\star\right)^{n_s-1}$, with no explicit dependence on the potential. Thus, \textsc{ModeCode} helps us to, from a given $V(\phi)$, solve the Klein-Gordon and Mukhanov-Sasaki equations, for the quantity $u\equiv -z\mathcal{R}$, with $\mathcal{R}$ being the curvature perturbation, $z\equiv\frac{a\dot\phi}{H}$, and $\mathcal{P}_\mathcal{R}$ is found more generically as $\mathcal{P}_\mathcal{R}=\frac{k^3}{2\pi^2}\left|\frac{u_k}{z}\right|^2$.
	
	Then, interfaced with the \textsc{CosmoMC} code \cite{Easther:2011yq}, we perform a Monte Carlo Markov Chain (MCMC) analysis of the model in order to constrain the values of the free parameters that fit the data best. It is also possible to compare two models and see which is more favoured by data, considering the models' complexity, such as the number of free parameters. Bayesian comparison is a common way of selecting models in such a manner \cite{Trotta:2008qt}. Models with many degrees of freedom will most certainly give a better fit to data at the cost of being disfavored by statistical criteria; therefore, generally, the key is to achieve a model that fits the data well without the need for many free parameters.

    \begin{table*}[ht]
		\centering
		\begin{tabular*}{\textwidth}{>{\footnotesize}c >{\footnotesize}c >{\footnotesize}c >{\footnotesize}c >{\footnotesize}c >{\footnotesize}c >{\footnotesize}c}
			\hline
			\hline
			& $\Lambda$CDM$+r$ & & NMWR (Fixed $M$) & & NMWR (Varying $M$) & \\
			\hline
			{Parameter} & {mean} & {best fit} & {mean} & {best fit}& {mean} & {best fit}\\
			\hline
			$\Omega_b h^2$   & $0.02219 \pm 0.00019$ & $0.0223$     & $0.0215\pm 0.00019$ & $0.0222$     & $0.02218\pm 0.00019$  & $0.0222$\\
			$\Omega_{c} h^2$ & $0.119\pm0.0011$ & $0.118$      & $0.120\pm 0.001$ & $0.120$      & $0.119\pm 0.0010$ &  $0.120$ \\	
			$100\theta$ & $1.041\pm0.00042$ & $1.041$     & $1.040\pm 0.00041$ & $1.040$      & $1.040\pm 0.00041 $  &  $1.041$ \\
			$\tau$ & $0.056\pm0.0072$ & $0.057$     & $0.05\pm 0.0026$ & $0.047$      & $0.050\pm 0.0037$ & $0.051$ \\
			$\log_{10}(M/M_p)$ & $-$ & $-$ & $-$ & $-$     & $-1.226\pm0.971$ & $-1.326$\\
			$\xi$ & $-$ & $-$ & $0.00093\pm 0.00036$ & $0.00085$     & $0.00098\pm 0.00032$ & $0.001$\\
			$H_{0}^{\ast}$ [Km/s/Mpc] & $67.49\pm0.5$ & $68.0$     & $67.31\pm 0.46$ & $67.18$      & $67.43\pm 0.47$ & $67.04$ \\
			$\Omega_{m}^{\ast}$ & $0.312\pm0.0067$ & $0.304$     & $0.314\pm 0.0062$ & $0.316$      & $0.312\pm0.0062$ & $0.318$\\
			$\Omega_{\Lambda}^{\ast}$ & $0.688\pm0.0067$ & $0.695$      & $0.685\pm 0.0062$ & $0.683$      & $0.687\pm 0.0062$ & $0.681$\\
			$n_s^{\ast}$ & $0.965\pm0.004$ & $0.9664$       & & $0.968$     & & $0.966$ \\
			$r_{0.002}^{\ast}$ & $<0.024$ & $0.013$       & $-$ & $0.036$     & $-$ & $0.013$ \\
            \hline
			$\Delta$DIC & Reference & & $-1.612$ & & $-1.628$ & \\
			\hline
		\end{tabular*}
		\caption{The estimates at $68\%$ confidence level (C.L.) and best-fit values for the cosmological parameters. The first columns show the constraints on the $\Lambda$CDM+r model, followed by the non-minimally coupled WR models, where we have considered the scenarios where $M=M_p$ and when the mass scale is free to vary.}
		\label{tab:2}
	\end{table*}

	In the present analysis, we consider the following cosmological parameters: The baryon and cold dark matter density parameters $\Omega_bh^2,\Omega_ch^2$, the ratio of the sound horizon and the angular distance $\theta$ at decoupling, the optical depth $\tau$, as well as the free parameters from the inflationary model $\xi$ and $M$, resulting in a total of six varying parameters \footnote{We set the amplitude of scalar perturbations $\mathcal{P}_\mathcal{R}$ at the Planck value, used to fix the amplitude of the potential $V_0$ from Eq. (\ref{eq:2.5}).}. We also fix the sum of neutrino masses to $0.06$ eV and use the pivot scale chosen by Planck, of $k_\star=0.05$ Mpc$^{-1}$, at $N_\star=55$. Table \ref{tab:1} shows the priors chosen for the analysis. We have chosen the interval of $\xi$ based on the behaviour of the $n_s$ and $r$ parameters; in particular, $n_s$ changes considerably, as for $\xi\simeq 0.002$ it starts to leave the Planck contours, so we take as our upper limit the value $\xi=0.003$, where we have also included the non-minimally coupled case $\xi=0$ as the lower one. As for $M$, we could, at first, decide to keep the same range as considered in a previous analysis of the minimally coupled model, performed in \cite{Martin:2013nzq}, especially because we note here that the presence of a non-zero $\xi$, allows for very high $M$ to be well inside the $1\sigma$ confidence contours, as seen in fig. \ref{fig:1}. However, our initial tests have shown that such high values are completely discarded by data,  For this reason, we restrict the analysis to a smaller range of $M$, of $\log_{10}M/M_p=[-3:2]$. We also perform an analysis based on the original setting of the model \cite{Albrecht:1983ib}, with $M$ fixed at $M_p$, to see how only the $\xi$ parameter affects the results.

    \begin{table}[ht]
		\begin{center}
        \begin{tabular}{ccccc}
			\hline
			\hline
			& WR (Fixed $M$) & & WR (Varying $M$) & \\
			\hline
			{Parameter} & {mean} & {best fit}& {mean} & {best fit}\\
			\hline
			$\Omega_b h^2$ & $0.02230\pm 0.00018$ & $0.02234$ & $0.0228\pm 0.00018$  & $0.0227$\\
			$\Omega_{c} h^2$  & $0.1181\pm 0.00087$ & $0.118$ & $0.1181\pm 0.00092$  & $0.119$\\
			$100\theta$  & $1.0411\pm 0.0004$ & $1.041$ & $1.0411\pm 0.00041$  & $1.041$\\
			$\tau$ & $0.0564\pm 0.0026$ & $0.0555$ & $0.0543\pm 0.004$  & $0.0545$\\
			$\log_{10}(M/M_p)$  & $-$ & $-$ & $-0.271\pm 0.8$  & $0.0775$\\
			$H_{0}^{\ast}$ [Km/s/Mpc] & $68.0\pm 0.385$ & $68.0$ & $68.0\pm 0.417$  & $67.65$\\
			$\Omega_{m}^{\ast}$ & $0.305\pm 0.005$ & $0.305$ & $0.304\pm 0.005$  & $0.310$\\
			$\Omega_{\Lambda}^{\ast}$  & $0.695\pm 0.005$ & $0.694$ & $0.695\pm 0.005$  & $0.689$\\
			$n_s^{\ast}$  & $-$ & $0.973$ & $-$  & $0.973$\\
			$r_{0.002}^{\ast}$ & $-$ & $0.048$ & $-$  & $0.05$\\
            \hline
			$\Delta$DIC & $0.86$ & & $3.642$ & \\
			\hline
		\end{tabular}
		\caption{Same as Tab. \ref{tab:2}, but now we consider the minimally-coupled WR model, again, both when $M=M_p$, and for a varying $M$.}
		\label{tab:3}
        \end{center}
	\end{table}
	
	For the data used in the analysis, we will work with CMB data from the latest Planck report \cite{Aghanim:2018eyx}, where we consider both high multipoles data from 100-,143-,217-GHz T maps and low multipoles from the TT (temperature), EE (E-mode polarization), BB (B-mode polarization) and TE (cross-correlation temperature polarization) likelihoods. We also combine the CMB data with BAO measurements from the 6dF Galaxy Survey (6dFGS) \cite{Beutler:2011hx}, the Sloan Digital Sky Survey (SDSS) \cite{Ross:2014qpa} and the BOSSgalaxy samples LOWZ and CMASS \cite{BOSS:2013rlg}.
	
	We also want to know whether this scenario is viable compared to the standard model. This information is obtained with statistical tools, such as information criteria \cite{Liddle:2007fy}. In our case, we choose the Deviance Information Criterion (DIC) as a means of comparison \cite{Spiegelhalter:2002yvw}. This quantity is constructed from the whole likelihood instead of using only the maximum value, being defined as
	\begin{gather}
		DIC_{\mathcal{M}}\equiv-2\overline{\operatorname{log}\mathcal{L}}(\theta) + p_D,
	\end{gather} 
	with the first term being the posterior mean of the likelihood $\mathcal{L}(\theta)$, given the parameters $\theta$, while the second term being the Bayesian complexity, $p_D=-2\operatorname{log}\overline{\mathcal{L}(\theta)} + 2\operatorname{log}\mathcal{L}(\bar\theta)$, so that the DIC is rewritten as DIC$_{\mathcal{M}}=-4\operatorname{log}\overline{\mathcal{L}(\theta)} + 2\operatorname{log}\mathcal{L}(\bar\theta)$, and since it accounts for the complexity of the model, more free parameters should generally penalize the model compared to a simpler one. Having computed the DIC for all models, one can compare them through the difference $\Delta$DIC=DIC$_\mathcal{M}$-DIC$_{ref}$, where the subscript $_{ref}$ corresponds to a reference model, that we take as being the $\Lambda$CDM one. Following \cite{dosSantos:2021vis,Winkler:2019hkh}, the data prefer the model with the lowest DIC, and in general, the scale $\Delta$DIC$=10/5/1$ corresponds to a strong/moderate/null preference for the reference model, respectively; negative values will then represent a preference for the proposed model. 
	
    \begin{figure*}[t]
			\includegraphics[width=\columnwidth]{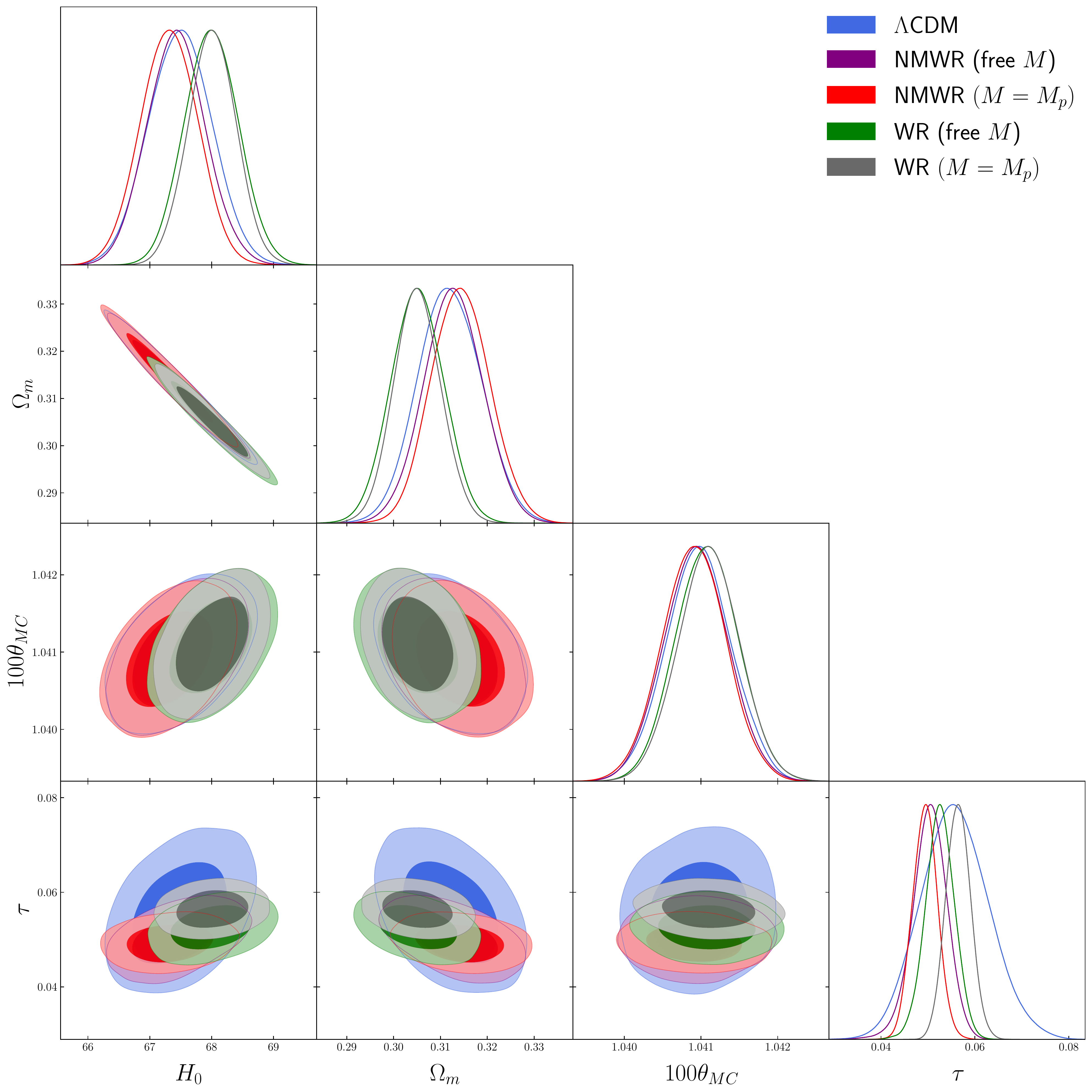}
		\caption{Confidence contours and posteriors for the models investigated in this work. The $\Lambda$CDM model is shown in blue, while the WR model is divided as the minimally-coupled one with $M=M_p$ (grey), varying $M$ (green) and non-minimally coupled, where the $M=M_p$ model is shown in red, while the varying $M$ one is shown in purple.}
		\label{fig:3}
	\end{figure*}

	\section{Results of the analysis}\label{sec4}

\begin{figure*}[t]
		    \includegraphics[width=0.5\columnwidth]{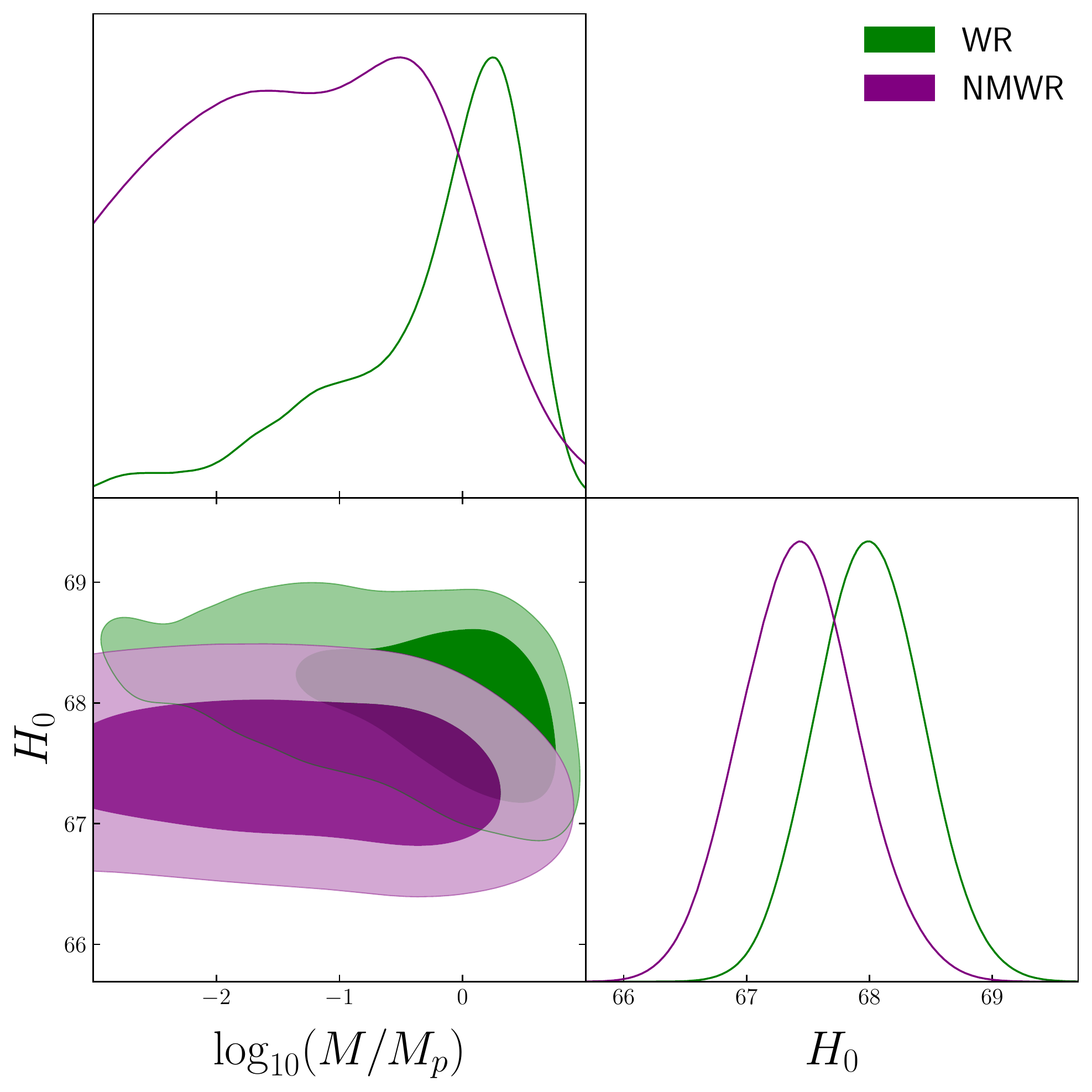}
			\includegraphics[width=0.5\columnwidth]{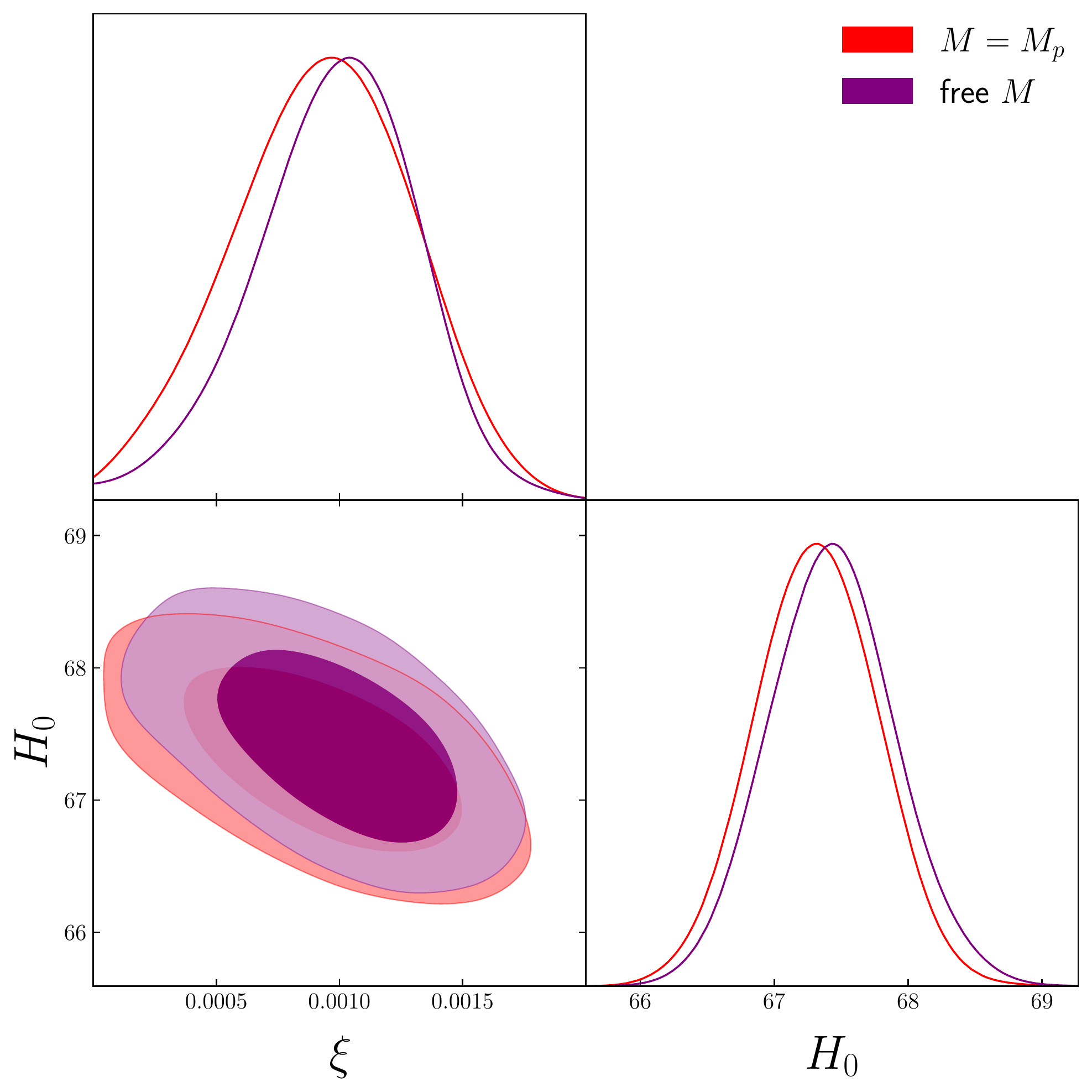}
		\caption{Triangle plots for the parameters of the WR model with the present Hubble parameter $H_0$. On the left panel, we show the $\log_{10}M/M_p-H_0$ plane, where the colours follow what is shown in Fig. \ref{fig:3}, while the right panel shows only the non-minimally coupled WR model, with the $\xi-H_0$ plane.}
		\label{fig:4}
	\end{figure*}

	In tables \ref{tab:2} and \ref{tab:3}, we show the results of the statistical analyses, with the values being at 68$\%$ confidence level. We start with the non-minimally coupled model, showing the results when the mass scale $M$ is fixed at $M=M_p$, with $\xi$ being the only free parameter of the potential. We find a value of $\xi=0.00093\pm0.00036$, corresponding to a pretty slight deviation from the minimally-coupled scenario; but, as we discussed, a slight change in $\xi$ has a significant impact on the predictions of the $n_s$ and $r$ parameters, seen in fig. \ref{fig:1}. Consequently, this value is enough for the model's predictions in the $n_s-r$ plane to agree with Planck's $1\sigma$ confidence region. The confidence contours for all the models are shown in figs. \ref{fig:3} and \ref{fig:4}, where we note that the non-minimally coupled model gives a lower value for the optical depth $\tau$. Also, in fig. \ref{fig:4} (right panel), we see that the minimally coupled limit is essentially excluded at $2\sigma$ confidence level, while the correlation between $\xi$ and $H_0$ shows that the present Hubble parameter decreases with growing $\xi$. We note that the other cosmological parameters are constrained to values much similar to those of the $\Lambda$CDM model, with a present value of the Hubble parameter of $H_0=67.31\pm0.46$ Km/s/Mpc, and matter density parameter of $\Omega_m=0.314\pm0.0062$. We then conclude that the non-minimally coupled WR model can reproduce the cosmological data very well, with the cost of one free parameter being the strength of the coupling of the field with gravity.  

    Next, we consider a varying $M$. We estimate $\log_{10}(M/M_p)=-1.226\pm 0.971$, corresponding to $M\simeq 0.059 M_p$, as it is noticeable a preference for a lower $M$ concerning the reduced Planck mass. It is shown in fig. \ref{fig:4} (left panel), where the upper limit is well-defined. On the right panel, we again note the anti-correlation between $\xi$ and $H_0$, also showing a preference for the presence of a non-minimal coupling within the model. The estimated value for $\xi$ makes the model even more concordant with the $n_s-r$ plane shown in fig. \ref{fig:1}, being $\xi=0.00098\pm 0.00032$.
	
\begin{figure*}[t]
    \centering
	\includegraphics[width=0.8\columnwidth]{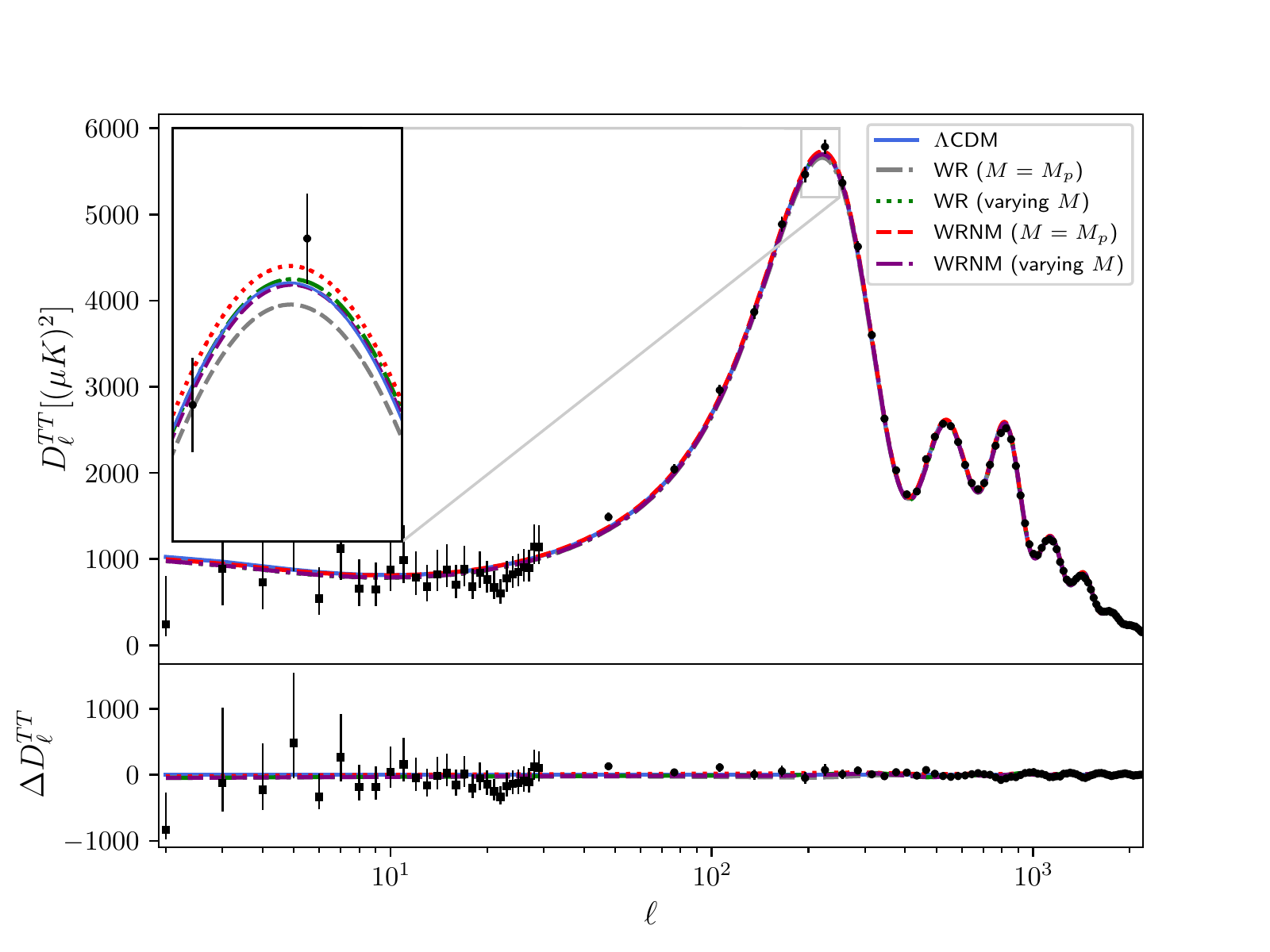}
	\caption{Best fit curves of the temperature power spectra for the models investigated, confronted with observational estimates from Planck \cite{Aghanim:2018eyx}.}
	\label{fig:5}
	\end{figure*}

	In tab. \ref{tab:3}, we show the results on the minimally-coupled WR model. We note that in both $M=M_p$ and varying $M$ cases, the estimated values of the cosmological parameters are almost identical, as in the free $M$ model, a mass close to the reduced Planck one is preferred, which is possible to see in fig. \ref{fig:4} (left panel), contrasting with the non-minimally coupled model. An interesting feature of the minimally-coupled model is that the $H_0$ estimate is the highest, being $H_0=68.0$ Km/s/Mpc, while the other cosmological parameters are more compatible with the $\Lambda$CDM model. We remember, however, that the concordance with the $n_s-r$ plane is not as good as in the non-minimally coupled model, as seen in fig. \ref{fig:1}. The best-fit curves for each model are shown in fig. \ref{fig:5}, where we note that the models where $M$ is varying give an excellent fit compared to the $\Lambda$CDM model. At the same time, a noticeable difference can be visualized for the minimally coupled model with $M=M_p$, which displays a lower amplitude of the first peak of the spectrum. Also, we see that the curves of the minimally-coupled models considered are different, although the results of the cosmological parameters are essentially the same. This is likely due to the slightly larger optical depth $\tau$ predicted by the $M=M_p$ case, which generally decreases the amplitude of the peaks.

	Finally, using the Deviance Information Criterion, we compare the WR model with the standard $\Lambda$CDM one. Starting from the minimally-coupled models, we note that regardless of a varying $M$, the WR model is shown to be disfavored concerning the standard one. For $M=M_p$, we still can consider the models to be statistically equivalent, but for a varying $M$, we obtain $\Delta$DIC=$3.642$, showing a moderate tension between models. The situation changes, however, when the non-minimally coupled model is considered. We have similar results for both fixed and varying $M$ models, where $\Delta$DIC=$-1.612$ and $\Delta$DIC=$-1.628$, respectively, display a weak preference for the non-minimally coupled WR model.

	\section{Discussion and Conclusions}\label{sec5}
	
	In this work, we investigated what the most recent CMB data tells about modifications of gravity in which a scalar field is non-minimally coupled to gravity, built to describe cosmic inflation. As an explicit example, we have considered the Witten-O'Raifeartaigh model, motivated through a supersymmetric formalism. The simplicity of the resulting function makes the model attractive from the phenomenological point of view, as the potential has only one parameter characterized by a mass scale that determines the minimum of the potential. Although a minimally-coupled version of the model predicts reasonable values for the spectral index $n_s$ and tensor-to-scalar ratio $r$, the more strict restrictions from CMB data has made concordance of its predictions more difficult, especially if we compare them with other models. This motivates the consideration of extended scenarios, particularly the presence of a non-minimal coupling of the inflaton with gravity. An initial analysis was made in \cite{Santos:2022aeb}, and it was found that a small value of the parameter $\xi$ was enough to drive the predictions of $n_s$ and $r$ to values well into the $1\sigma$ confidence region of Planck data. Moreover, the range of the non-minimal coupling parameter is shown to be very restricted, in a way that it becomes interesting to know whether the CMB data indicates the presence of a non-zero $\xi$, at the same time that one can check if the value is consistent with the $n_s-r$ plane.

	While obtaining new constraints on both minimally and non-minimally coupled WR models, we have considered two regimes for each one of them: Fixing the mass scale as $M=M_p$ has to do with the original motivation of the model in which the mass scale $M$ is theoretically motivated to be the order of Planck mass; we then only seek to constrain $\xi$, from which we find a preference for $\xi>0$ at $2\sigma$ confidence level. Although the value is small, as discussed, it is enough to solve the tension with the $1\sigma$ region of the $n_s-r$ plane present in the minimally-coupled model. Allowing $M$ to vary also leads to interesting results. A higher value of $\xi$ is found, of $\xi\sim 0.001$, driving the predictions of the $n_s-r$ plane to the $1\sigma$ confidence region. As for the mass scale $M$ itself, we find that while there was some difficulty in establishing a precise estimate for the parameter, a value at least less than $M_p$ seems to be preferred, so an upper limit is well defined. Finally, we compare the models with the standard $\Lambda$CDM using the deviance information criterion. Regardless of varying $M$, a weak preference for the WR model is obtained, motivating further investigation of extended gravity scenarios, especially in light of future cosmic microwave background observations \cite{Hazumi:2019lys}. Moreover, it can be interesting to investigate such a scenario in a complete setup where preheating effects take place, as the interval of $M$ constrained follows the resonant tachyonic amplification of fluctuations of the field, leading to particle production and possible generation of primordial gravitational waves \cite{Felder:2000hj,Felder:2001kt,Garcia-Bellido:2007fiu,Antusch:2016con}. All these questions are to be addressed in future works.

	\section*{Acknowledgments}
	
	F.B.M. dos Santos is supported by Coordenação de Aperfeiçoamento de Pessoal de
	Nível Superior (CAPES). R. Silva acknowledges financial support from CNPq (Grant
	No. 307620/2019-0). JSA is supported by CNPq (Grants no. 310790/2014-0 and 400471/2014-0) and Funda\c{c}\~ao de Amparo \`a Pesquisa do Estado do Rio de Janeiro FAPERJ (grant no. 233906). We also acknowledge the use of CosmoMC and ModeCode packages. This work was developed thanks to the High-Performance Computing Center at the Universidade Federal do Rio Grande do Norte (NPAD/UFRN).
	
	\bibliographystyle{ieeetr}
	\bibliography{references}

\begin{thebibliography}{53}%
\makeatletter
\providecommand \@ifxundefined [1]{%
 \@ifx{#1\undefined}
}%
\providecommand \@ifnum [1]{%
 \ifnum #1\expandafter \@firstoftwo
 \else \expandafter \@secondoftwo
 \fi
}%
\providecommand \@ifx [1]{%
 \ifx #1\expandafter \@firstoftwo
 \else \expandafter \@secondoftwo
 \fi
}%
\providecommand \natexlab [1]{#1}%
\providecommand \enquote  [1]{``#1''}%
\providecommand \bibnamefont  [1]{#1}%
\providecommand \bibfnamefont [1]{#1}%
\providecommand \citenamefont [1]{#1}%
\providecommand \href@noop [0]{\@secondoftwo}%
\providecommand \href [0]{\begingroup \@sanitize@url \@href}%
\providecommand \@href[1]{\@@startlink{#1}\@@href}%
\providecommand \@@href[1]{\endgroup#1\@@endlink}%
\providecommand \@sanitize@url [0]{\catcode `\\12\catcode `\$12\catcode
  `\&12\catcode `\#12\catcode `\^12\catcode `\_12\catcode `\%12\relax}%
\providecommand \@@startlink[1]{}%
\providecommand \@@endlink[0]{}%
\providecommand \url  [0]{\begingroup\@sanitize@url \@url }%
\providecommand \@url [1]{\endgroup\@href {#1}{\urlprefix }}%
\providecommand \urlprefix  [0]{URL }%
\providecommand \Eprint [0]{\href }%
\providecommand \doibase [0]{http://dx.doi.org/}%
\providecommand \selectlanguage [0]{\@gobble}%
\providecommand \bibinfo  [0]{\@secondoftwo}%
\providecommand \bibfield  [0]{\@secondoftwo}%
\providecommand \translation [1]{[#1]}%
\providecommand \BibitemOpen [0]{}%
\providecommand \bibitemStop [0]{}%
\providecommand \bibitemNoStop [0]{.\EOS\space}%
\providecommand \EOS [0]{\spacefactor3000\relax}%
\providecommand \BibitemShut  [1]{\csname bibitem#1\endcsname}%
\let\auto@bib@innerbib\@empty
\bibitem [{\citenamefont {Guth}(1981)}]{Guth1981}%
  \BibitemOpen
  \bibfield  {author} {\bibinfo {author} {\bibfnamefont {A.~H.}\ \bibnamefont
  {Guth}},\ }\href {\doibase 10.1103/physrevd.23.347} {\bibfield  {journal}
  {\bibinfo  {journal} {Physical Review D}\ }\textbf {\bibinfo {volume} {23}},\
  \bibinfo {pages} {347} (\bibinfo {year} {1981})}\BibitemShut {NoStop}%
\bibitem [{\citenamefont {Starobinsky}(1980)}]{Starobinsky:1980te}%
  \BibitemOpen
  \bibfield  {author} {\bibinfo {author} {\bibfnamefont {A.~A.}\ \bibnamefont
  {Starobinsky}},\ }\href {\doibase 10.1016/0370-2693(80)90670-X} {\bibfield
  {journal} {\bibinfo  {journal} {Phys. Lett. B}\ }\textbf {\bibinfo {volume}
  {91}},\ \bibinfo {pages} {99} (\bibinfo {year} {1980})}\BibitemShut {NoStop}%
\bibitem [{\citenamefont {Linde}(1982)}]{Linde:1981mu}%
  \BibitemOpen
  \bibfield  {author} {\bibinfo {author} {\bibfnamefont {A.~D.}\ \bibnamefont
  {Linde}},\ }\href {\doibase 10.1016/0370-2693(82)91219-9} {\bibfield
  {journal} {\bibinfo  {journal} {Phys. Lett. B}\ }\textbf {\bibinfo {volume}
  {108}},\ \bibinfo {pages} {389} (\bibinfo {year} {1982})}\BibitemShut
  {NoStop}%
\bibitem [{\citenamefont {Linde}(1983)}]{Linde:1983gd}%
  \BibitemOpen
  \bibfield  {author} {\bibinfo {author} {\bibfnamefont {A.~D.}\ \bibnamefont
  {Linde}},\ }\href {\doibase 10.1016/0370-2693(83)90837-7} {\bibfield
  {journal} {\bibinfo  {journal} {Phys. Lett. B}\ }\textbf {\bibinfo {volume}
  {129}},\ \bibinfo {pages} {177} (\bibinfo {year} {1983})}\BibitemShut
  {NoStop}%
\bibitem [{\citenamefont {Mather}\ \emph {et~al.}(1994)\citenamefont {Mather}
  \emph {et~al.}}]{Mather:1993ij}%
  \BibitemOpen
  \bibfield  {author} {\bibinfo {author} {\bibfnamefont {J.~C.}\ \bibnamefont
  {Mather}} \emph {et~al.},\ }\href {\doibase 10.1086/173574} {\bibfield
  {journal} {\bibinfo  {journal} {Astrophys. J.}\ }\textbf {\bibinfo {volume}
  {420}},\ \bibinfo {pages} {439} (\bibinfo {year} {1994})}\BibitemShut
  {NoStop}%
\bibitem [{\citenamefont {Peiris}\ \emph {et~al.}(2003)\citenamefont {Peiris}
  \emph {et~al.}}]{WMAP:2003syu}%
  \BibitemOpen
  \bibfield  {author} {\bibinfo {author} {\bibfnamefont {H.~V.}\ \bibnamefont
  {Peiris}} \emph {et~al.} (\bibinfo {collaboration} {WMAP}),\ }\href {\doibase
  10.1086/377228} {\bibfield  {journal} {\bibinfo  {journal} {Astrophys. J.
  Suppl.}\ }\textbf {\bibinfo {volume} {148}},\ \bibinfo {pages} {213}
  (\bibinfo {year} {2003})},\ \Eprint {http://arxiv.org/abs/astro-ph/0302225}
  {arXiv:astro-ph/0302225} \BibitemShut {NoStop}%
\bibitem [{\citenamefont {Spergel}\ \emph {et~al.}(2007)\citenamefont {Spergel}
  \emph {et~al.}}]{WMAP:2006bqn}%
  \BibitemOpen
  \bibfield  {author} {\bibinfo {author} {\bibfnamefont {D.~N.}\ \bibnamefont
  {Spergel}} \emph {et~al.} (\bibinfo {collaboration} {WMAP}),\ }\href
  {\doibase 10.1086/513700} {\bibfield  {journal} {\bibinfo  {journal}
  {Astrophys. J. Suppl.}\ }\textbf {\bibinfo {volume} {170}},\ \bibinfo {pages}
  {377} (\bibinfo {year} {2007})},\ \Eprint
  {http://arxiv.org/abs/astro-ph/0603449} {arXiv:astro-ph/0603449} \BibitemShut
  {NoStop}%
\bibitem [{\citenamefont {Aghanim}\ \emph {et~al.}(2020)\citenamefont {Aghanim}
  \emph {et~al.}}]{Aghanim:2018eyx}%
  \BibitemOpen
  \bibfield  {author} {\bibinfo {author} {\bibfnamefont {N.}~\bibnamefont
  {Aghanim}} \emph {et~al.} (\bibinfo {collaboration} {Planck}),\ }\href
  {\doibase 10.1051/0004-6361/201833910} {\bibfield  {journal} {\bibinfo
  {journal} {Astron. Astrophys.}\ }\textbf {\bibinfo {volume} {641}},\ \bibinfo
  {pages} {A6} (\bibinfo {year} {2020})},\ \Eprint
  {http://arxiv.org/abs/1807.06209} {arXiv:1807.06209 [astro-ph.CO]}
  \BibitemShut {NoStop}%
\bibitem [{\citenamefont {Abbott}\ \emph {et~al.}(1982)\citenamefont {Abbott},
  \citenamefont {Farhi},\ and\ \citenamefont {Wise}}]{Abbott:1982hn}%
  \BibitemOpen
  \bibfield  {author} {\bibinfo {author} {\bibfnamefont {L.~F.}\ \bibnamefont
  {Abbott}}, \bibinfo {author} {\bibfnamefont {E.}~\bibnamefont {Farhi}}, \
  and\ \bibinfo {author} {\bibfnamefont {M.~B.}\ \bibnamefont {Wise}},\ }\href
  {\doibase 10.1016/0370-2693(82)90867-X} {\bibfield  {journal} {\bibinfo
  {journal} {Phys. Lett. B}\ }\textbf {\bibinfo {volume} {117}},\ \bibinfo
  {pages} {29} (\bibinfo {year} {1982})}\BibitemShut {NoStop}%
\bibitem [{\citenamefont {Albrecht}\ \emph {et~al.}(1982)\citenamefont
  {Albrecht}, \citenamefont {Steinhardt}, \citenamefont {Turner},\ and\
  \citenamefont {Wilczek}}]{Albrecht:1982mp}%
  \BibitemOpen
  \bibfield  {author} {\bibinfo {author} {\bibfnamefont {A.}~\bibnamefont
  {Albrecht}}, \bibinfo {author} {\bibfnamefont {P.~J.}\ \bibnamefont
  {Steinhardt}}, \bibinfo {author} {\bibfnamefont {M.~S.}\ \bibnamefont
  {Turner}}, \ and\ \bibinfo {author} {\bibfnamefont {F.}~\bibnamefont
  {Wilczek}},\ }\href {\doibase 10.1103/PhysRevLett.48.1437} {\bibfield
  {journal} {\bibinfo  {journal} {Phys. Rev. Lett.}\ }\textbf {\bibinfo
  {volume} {48}},\ \bibinfo {pages} {1437} (\bibinfo {year}
  {1982})}\BibitemShut {NoStop}%
\bibitem [{\citenamefont {Kofman}\ \emph {et~al.}(1994)\citenamefont {Kofman},
  \citenamefont {Linde},\ and\ \citenamefont {Starobinsky}}]{Kofman:1994rk}%
  \BibitemOpen
  \bibfield  {author} {\bibinfo {author} {\bibfnamefont {L.}~\bibnamefont
  {Kofman}}, \bibinfo {author} {\bibfnamefont {A.~D.}\ \bibnamefont {Linde}}, \
  and\ \bibinfo {author} {\bibfnamefont {A.~A.}\ \bibnamefont {Starobinsky}},\
  }\href {\doibase 10.1103/PhysRevLett.73.3195} {\bibfield  {journal} {\bibinfo
   {journal} {Phys. Rev. Lett.}\ }\textbf {\bibinfo {volume} {73}},\ \bibinfo
  {pages} {3195} (\bibinfo {year} {1994})},\ \Eprint
  {http://arxiv.org/abs/hep-th/9405187} {arXiv:hep-th/9405187} \BibitemShut
  {NoStop}%
\bibitem [{\citenamefont {Martin}\ \emph
  {et~al.}(2014{\natexlab{a}})\citenamefont {Martin}, \citenamefont
  {Ringeval},\ and\ \citenamefont {Vennin}}]{Martin:2013tda}%
  \BibitemOpen
  \bibfield  {author} {\bibinfo {author} {\bibfnamefont {J.}~\bibnamefont
  {Martin}}, \bibinfo {author} {\bibfnamefont {C.}~\bibnamefont {Ringeval}}, \
  and\ \bibinfo {author} {\bibfnamefont {V.}~\bibnamefont {Vennin}},\ }\href
  {\doibase 10.1016/j.dark.2014.01.003} {\bibfield  {journal} {\bibinfo
  {journal} {Phys. Dark Univ.}\ }\textbf {\bibinfo {volume} {5-6}},\ \bibinfo
  {pages} {75} (\bibinfo {year} {2014}{\natexlab{a}})},\ \Eprint
  {http://arxiv.org/abs/1303.3787} {arXiv:1303.3787 [astro-ph.CO]} \BibitemShut
  {NoStop}%
\bibitem [{\citenamefont {Martin}\ \emph
  {et~al.}(2014{\natexlab{b}})\citenamefont {Martin}, \citenamefont {Ringeval},
  \citenamefont {Trotta},\ and\ \citenamefont {Vennin}}]{Martin:2013nzq}%
  \BibitemOpen
  \bibfield  {author} {\bibinfo {author} {\bibfnamefont {J.}~\bibnamefont
  {Martin}}, \bibinfo {author} {\bibfnamefont {C.}~\bibnamefont {Ringeval}},
  \bibinfo {author} {\bibfnamefont {R.}~\bibnamefont {Trotta}}, \ and\ \bibinfo
  {author} {\bibfnamefont {V.}~\bibnamefont {Vennin}},\ }\href {\doibase
  10.1088/1475-7516/2014/03/039} {\bibfield  {journal} {\bibinfo  {journal}
  {JCAP}\ }\textbf {\bibinfo {volume} {03}},\ \bibinfo {pages} {039} (\bibinfo
  {year} {2014}{\natexlab{b}})},\ \Eprint {http://arxiv.org/abs/1312.3529}
  {arXiv:1312.3529 [astro-ph.CO]} \BibitemShut {NoStop}%
\bibitem [{\citenamefont {Lucchin}\ \emph {et~al.}(1986)\citenamefont
  {Lucchin}, \citenamefont {Matarrese},\ and\ \citenamefont
  {Pollock}}]{Lucchin:1985ip}%
  \BibitemOpen
  \bibfield  {author} {\bibinfo {author} {\bibfnamefont {F.}~\bibnamefont
  {Lucchin}}, \bibinfo {author} {\bibfnamefont {S.}~\bibnamefont {Matarrese}},
  \ and\ \bibinfo {author} {\bibfnamefont {M.~D.}\ \bibnamefont {Pollock}},\
  }\href {\doibase 10.1016/0370-2693(86)90592-7} {\bibfield  {journal}
  {\bibinfo  {journal} {Phys. Lett. B}\ }\textbf {\bibinfo {volume} {167}},\
  \bibinfo {pages} {163} (\bibinfo {year} {1986})}\BibitemShut {NoStop}%
\bibitem [{\citenamefont {Futamase}\ and\ \citenamefont
  {Maeda}(1989)}]{Futamase:1987ua}%
  \BibitemOpen
  \bibfield  {author} {\bibinfo {author} {\bibfnamefont {T.}~\bibnamefont
  {Futamase}}\ and\ \bibinfo {author} {\bibfnamefont {K.-i.}\ \bibnamefont
  {Maeda}},\ }\href {\doibase 10.1103/PhysRevD.39.399} {\bibfield  {journal}
  {\bibinfo  {journal} {Phys. Rev. D}\ }\textbf {\bibinfo {volume} {39}},\
  \bibinfo {pages} {399} (\bibinfo {year} {1989})}\BibitemShut {NoStop}%
\bibitem [{\citenamefont {Fakir}\ and\ \citenamefont
  {Unruh}(1990)}]{Fakir1990}%
  \BibitemOpen
  \bibfield  {author} {\bibinfo {author} {\bibfnamefont {R.}~\bibnamefont
  {Fakir}}\ and\ \bibinfo {author} {\bibfnamefont {W.~G.}\ \bibnamefont
  {Unruh}},\ }\href {\doibase 10.1103/physrevd.41.1783} {\bibfield  {journal}
  {\bibinfo  {journal} {Physical Review D}\ }\textbf {\bibinfo {volume} {41}},\
  \bibinfo {pages} {1783} (\bibinfo {year} {1990})}\BibitemShut {NoStop}%
\bibitem [{\citenamefont {Komatsu}\ and\ \citenamefont
  {Futamase}(1998)}]{Komatsu1998}%
  \BibitemOpen
  \bibfield  {author} {\bibinfo {author} {\bibfnamefont {E.}~\bibnamefont
  {Komatsu}}\ and\ \bibinfo {author} {\bibfnamefont {T.}~\bibnamefont
  {Futamase}},\ }\href {\doibase 10.1103/physrevd.58.023004} {\bibfield
  {journal} {\bibinfo  {journal} {Physical Review D}\ }\textbf {\bibinfo
  {volume} {58}} (\bibinfo {year} {1998}),\
  10.1103/physrevd.58.023004}\BibitemShut {NoStop}%
\bibitem [{\citenamefont {Komatsu}\ and\ \citenamefont
  {Futamase}(1999)}]{Komatsu1999}%
  \BibitemOpen
  \bibfield  {author} {\bibinfo {author} {\bibfnamefont {E.}~\bibnamefont
  {Komatsu}}\ and\ \bibinfo {author} {\bibfnamefont {T.}~\bibnamefont
  {Futamase}},\ }\href {\doibase 10.1103/physrevd.59.064029} {\bibfield
  {journal} {\bibinfo  {journal} {Physical Review D}\ }\textbf {\bibinfo
  {volume} {59}} (\bibinfo {year} {1999}),\
  10.1103/physrevd.59.064029}\BibitemShut {NoStop}%
\bibitem [{\citenamefont {Linde}\ \emph {et~al.}(2011)\citenamefont {Linde},
  \citenamefont {Noorbala},\ and\ \citenamefont {Westphal}}]{Linde:2011nh}%
  \BibitemOpen
  \bibfield  {author} {\bibinfo {author} {\bibfnamefont {A.}~\bibnamefont
  {Linde}}, \bibinfo {author} {\bibfnamefont {M.}~\bibnamefont {Noorbala}}, \
  and\ \bibinfo {author} {\bibfnamefont {A.}~\bibnamefont {Westphal}},\ }\href
  {\doibase 10.1088/1475-7516/2011/03/013} {\bibfield  {journal} {\bibinfo
  {journal} {JCAP}\ }\textbf {\bibinfo {volume} {03}},\ \bibinfo {pages} {013}
  (\bibinfo {year} {2011})},\ \Eprint {http://arxiv.org/abs/1101.2652}
  {arXiv:1101.2652 [hep-th]} \BibitemShut {NoStop}%
\bibitem [{\citenamefont {Hertzberg}(2010)}]{Hertzberg:2010dc}%
  \BibitemOpen
  \bibfield  {author} {\bibinfo {author} {\bibfnamefont {M.~P.}\ \bibnamefont
  {Hertzberg}},\ }\href {\doibase 10.1007/JHEP11(2010)023} {\bibfield
  {journal} {\bibinfo  {journal} {JHEP}\ }\textbf {\bibinfo {volume} {11}},\
  \bibinfo {pages} {023} (\bibinfo {year} {2010})},\ \Eprint
  {http://arxiv.org/abs/1002.2995} {arXiv:1002.2995 [hep-ph]} \BibitemShut
  {NoStop}%
\bibitem [{\citenamefont {Okada}\ \emph {et~al.}(2016)\citenamefont {Okada},
  \citenamefont {\c{S}eno\u{g}uz},\ and\ \citenamefont
  {Shafi}}]{Okada:2014lxa}%
  \BibitemOpen
  \bibfield  {author} {\bibinfo {author} {\bibfnamefont {N.}~\bibnamefont
  {Okada}}, \bibinfo {author} {\bibfnamefont {V.~N.}\ \bibnamefont
  {\c{S}eno\u{g}uz}}, \ and\ \bibinfo {author} {\bibfnamefont {Q.}~\bibnamefont
  {Shafi}},\ }\href {\doibase 10.3906/fiz-1505-7} {\bibfield  {journal}
  {\bibinfo  {journal} {Turk. J. Phys.}\ }\textbf {\bibinfo {volume} {40}},\
  \bibinfo {pages} {150} (\bibinfo {year} {2016})},\ \Eprint
  {http://arxiv.org/abs/1403.6403} {arXiv:1403.6403 [hep-ph]} \BibitemShut
  {NoStop}%
\bibitem [{\citenamefont {Kaiser}(2016)}]{Kaiser:2015usz}%
  \BibitemOpen
  \bibfield  {author} {\bibinfo {author} {\bibfnamefont {D.~I.}\ \bibnamefont
  {Kaiser}},\ }\href {\doibase 10.1007/978-3-319-31299-6_2} {\bibfield
  {journal} {\bibinfo  {journal} {Fundam. Theor. Phys.}\ }\textbf {\bibinfo
  {volume} {183}},\ \bibinfo {pages} {41} (\bibinfo {year} {2016})},\ \Eprint
  {http://arxiv.org/abs/1511.09148} {arXiv:1511.09148 [astro-ph.CO]}
  \BibitemShut {NoStop}%
\bibitem [{\citenamefont {Bezrukov}\ and\ \citenamefont
  {Shaposhnikov}(2008)}]{Bezrukov2008}%
  \BibitemOpen
  \bibfield  {author} {\bibinfo {author} {\bibfnamefont {F.}~\bibnamefont
  {Bezrukov}}\ and\ \bibinfo {author} {\bibfnamefont {M.}~\bibnamefont
  {Shaposhnikov}},\ }\href {\doibase 10.1016/j.physletb.2007.11.072} {\bibfield
   {journal} {\bibinfo  {journal} {Physics Letters B}\ }\textbf {\bibinfo
  {volume} {659}},\ \bibinfo {pages} {703} (\bibinfo {year}
  {2008})}\BibitemShut {NoStop}%
\bibitem [{\citenamefont {Tenkanen}(2017)}]{Tenkanen:2017jih}%
  \BibitemOpen
  \bibfield  {author} {\bibinfo {author} {\bibfnamefont {T.}~\bibnamefont
  {Tenkanen}},\ }\href {\doibase 10.1088/1475-7516/2017/12/001} {\bibfield
  {journal} {\bibinfo  {journal} {JCAP}\ }\textbf {\bibinfo {volume} {12}},\
  \bibinfo {pages} {001} (\bibinfo {year} {2017})},\ \Eprint
  {http://arxiv.org/abs/1710.02758} {arXiv:1710.02758 [astro-ph.CO]}
  \BibitemShut {NoStop}%
\bibitem [{\citenamefont {Campista}\ \emph {et~al.}(2017)\citenamefont
  {Campista}, \citenamefont {Benetti},\ and\ \citenamefont
  {Alcaniz}}]{Campista:2017ovq}%
  \BibitemOpen
  \bibfield  {author} {\bibinfo {author} {\bibfnamefont {M.}~\bibnamefont
  {Campista}}, \bibinfo {author} {\bibfnamefont {M.}~\bibnamefont {Benetti}}, \
  and\ \bibinfo {author} {\bibfnamefont {J.}~\bibnamefont {Alcaniz}},\ }\href
  {\doibase 10.1088/1475-7516/2017/09/010} {\bibfield  {journal} {\bibinfo
  {journal} {JCAP}\ }\textbf {\bibinfo {volume} {09}},\ \bibinfo {pages} {010}
  (\bibinfo {year} {2017})},\ \Eprint {http://arxiv.org/abs/1705.08877}
  {arXiv:1705.08877 [astro-ph.CO]} \BibitemShut {NoStop}%
\bibitem [{\citenamefont {Ferreira}\ \emph {et~al.}(2018)\citenamefont
  {Ferreira}, \citenamefont {Notari},\ and\ \citenamefont
  {Simeon}}]{Ferreira:2018nav}%
  \BibitemOpen
  \bibfield  {author} {\bibinfo {author} {\bibfnamefont {R.~Z.}\ \bibnamefont
  {Ferreira}}, \bibinfo {author} {\bibfnamefont {A.}~\bibnamefont {Notari}}, \
  and\ \bibinfo {author} {\bibfnamefont {G.}~\bibnamefont {Simeon}},\ }\href
  {\doibase 10.1088/1475-7516/2018/11/021} {\bibfield  {journal} {\bibinfo
  {journal} {JCAP}\ }\textbf {\bibinfo {volume} {11}},\ \bibinfo {pages} {021}
  (\bibinfo {year} {2018})},\ \Eprint {http://arxiv.org/abs/1806.05511}
  {arXiv:1806.05511 [astro-ph.CO]} \BibitemShut {NoStop}%
\bibitem [{\citenamefont {Bostan}\ \emph {et~al.}(2018)\citenamefont {Bostan},
  \citenamefont {G\"ulery\"uz},\ and\ \citenamefont
  {\c{S}eno\u{g}uz}}]{Bostan:2018evz}%
  \BibitemOpen
  \bibfield  {author} {\bibinfo {author} {\bibfnamefont {N.}~\bibnamefont
  {Bostan}}, \bibinfo {author} {\bibfnamefont {O.}~\bibnamefont
  {G\"ulery\"uz}}, \ and\ \bibinfo {author} {\bibfnamefont {V.~N.}\
  \bibnamefont {\c{S}eno\u{g}uz}},\ }\href {\doibase
  10.1088/1475-7516/2018/05/046} {\bibfield  {journal} {\bibinfo  {journal}
  {JCAP}\ }\textbf {\bibinfo {volume} {05}},\ \bibinfo {pages} {046} (\bibinfo
  {year} {2018})},\ \Eprint {http://arxiv.org/abs/1802.04160} {arXiv:1802.04160
  [astro-ph.CO]} \BibitemShut {NoStop}%
\bibitem [{\citenamefont {Reyimuaji}\ and\ \citenamefont
  {Zhang}(2021)}]{Reyimuaji:2020goi}%
  \BibitemOpen
  \bibfield  {author} {\bibinfo {author} {\bibfnamefont {Y.}~\bibnamefont
  {Reyimuaji}}\ and\ \bibinfo {author} {\bibfnamefont {X.}~\bibnamefont
  {Zhang}},\ }\href {\doibase 10.1088/1475-7516/2021/03/059} {\bibfield
  {journal} {\bibinfo  {journal} {JCAP}\ }\textbf {\bibinfo {volume} {03}},\
  \bibinfo {pages} {059} (\bibinfo {year} {2021})},\ \Eprint
  {http://arxiv.org/abs/2012.14248} {arXiv:2012.14248 [astro-ph.CO]}
  \BibitemShut {NoStop}%
\bibitem [{\citenamefont {Rodrigues}\ \emph {et~al.}(2021)\citenamefont
  {Rodrigues}, \citenamefont {Benetti},\ and\ \citenamefont
  {Alcaniz}}]{Rodrigues:2021txa}%
  \BibitemOpen
  \bibfield  {author} {\bibinfo {author} {\bibfnamefont {J.~G.}\ \bibnamefont
  {Rodrigues}}, \bibinfo {author} {\bibfnamefont {M.}~\bibnamefont {Benetti}},
  \ and\ \bibinfo {author} {\bibfnamefont {J.~S.}\ \bibnamefont {Alcaniz}},\
  }\href@noop {} {\  (\bibinfo {year} {2021})},\ \Eprint
  {http://arxiv.org/abs/2105.07009} {arXiv:2105.07009 [hep-ph]} \BibitemShut
  {NoStop}%
\bibitem [{\citenamefont {Rodrigues}\ \emph {et~al.}(2020)\citenamefont
  {Rodrigues}, \citenamefont {Benetti}, \citenamefont {Campista},\ and\
  \citenamefont {Alcaniz}}]{Rodrigues:2020dod}%
  \BibitemOpen
  \bibfield  {author} {\bibinfo {author} {\bibfnamefont {J.~G.}\ \bibnamefont
  {Rodrigues}}, \bibinfo {author} {\bibfnamefont {M.}~\bibnamefont {Benetti}},
  \bibinfo {author} {\bibfnamefont {M.}~\bibnamefont {Campista}}, \ and\
  \bibinfo {author} {\bibfnamefont {J.}~\bibnamefont {Alcaniz}},\ }\href
  {\doibase 10.1088/1475-7516/2020/07/007} {\bibfield  {journal} {\bibinfo
  {journal} {JCAP}\ }\textbf {\bibinfo {volume} {07}},\ \bibinfo {pages} {007}
  (\bibinfo {year} {2020})},\ \Eprint {http://arxiv.org/abs/2002.05154}
  {arXiv:2002.05154 [astro-ph.CO]} \BibitemShut {NoStop}%
\bibitem [{\citenamefont {dos Santos}\ \emph {et~al.}(2022)\citenamefont {dos
  Santos}, \citenamefont {Santos~da Costa}, \citenamefont {Silva},
  \citenamefont {Benetti},\ and\ \citenamefont {Alcaniz}}]{dosSantos:2021vis}%
  \BibitemOpen
  \bibfield  {author} {\bibinfo {author} {\bibfnamefont {F.~B.~M.}\
  \bibnamefont {dos Santos}}, \bibinfo {author} {\bibfnamefont
  {S.}~\bibnamefont {Santos~da Costa}}, \bibinfo {author} {\bibfnamefont
  {R.}~\bibnamefont {Silva}}, \bibinfo {author} {\bibfnamefont
  {M.}~\bibnamefont {Benetti}}, \ and\ \bibinfo {author} {\bibfnamefont
  {J.}~\bibnamefont {Alcaniz}},\ }\href {\doibase
  10.1088/1475-7516/2022/06/001} {\bibfield  {journal} {\bibinfo  {journal}
  {JCAP}\ }\textbf {\bibinfo {volume} {06}},\ \bibinfo {pages} {001} (\bibinfo
  {year} {2022})},\ \Eprint {http://arxiv.org/abs/2110.14758} {arXiv:2110.14758
  [astro-ph.CO]} \BibitemShut {NoStop}%
\bibitem [{\citenamefont {Santos}\ and\ \citenamefont
  {Silva}(2022)}]{Santos:2022aeb}%
  \BibitemOpen
  \bibfield  {author} {\bibinfo {author} {\bibfnamefont {F.~B. M.~d.}\
  \bibnamefont {Santos}}\ and\ \bibinfo {author} {\bibfnamefont
  {R.}~\bibnamefont {Silva}},\ }\href {\doibase 10.1088/1475-7516/2022/08/002}
  {\bibfield  {journal} {\bibinfo  {journal} {JCAP}\ }\textbf {\bibinfo
  {volume} {08}},\ \bibinfo {pages} {002} (\bibinfo {year} {2022})},\ \Eprint
  {http://arxiv.org/abs/2204.13694} {arXiv:2204.13694 [astro-ph.CO]}
  \BibitemShut {NoStop}%
\bibitem [{\citenamefont {Albrecht}\ \emph {et~al.}(1983)\citenamefont
  {Albrecht}, \citenamefont {Dimopoulos}, \citenamefont {Fischler},
  \citenamefont {Kolb}, \citenamefont {Raby},\ and\ \citenamefont
  {Steinhardt}}]{Albrecht:1983ib}%
  \BibitemOpen
  \bibfield  {author} {\bibinfo {author} {\bibfnamefont {A.}~\bibnamefont
  {Albrecht}}, \bibinfo {author} {\bibfnamefont {S.}~\bibnamefont
  {Dimopoulos}}, \bibinfo {author} {\bibfnamefont {W.}~\bibnamefont
  {Fischler}}, \bibinfo {author} {\bibfnamefont {E.~W.}\ \bibnamefont {Kolb}},
  \bibinfo {author} {\bibfnamefont {S.}~\bibnamefont {Raby}}, \ and\ \bibinfo
  {author} {\bibfnamefont {P.~J.}\ \bibnamefont {Steinhardt}},\ }\href
  {\doibase 10.1016/0550-3213(83)90347-4} {\bibfield  {journal} {\bibinfo
  {journal} {Nucl. Phys. B}\ }\textbf {\bibinfo {volume} {229}},\ \bibinfo
  {pages} {528} (\bibinfo {year} {1983})}\BibitemShut {NoStop}%
\bibitem [{\citenamefont {Witten}(1981{\natexlab{a}})}]{Witten:1981kv}%
  \BibitemOpen
  \bibfield  {author} {\bibinfo {author} {\bibfnamefont {E.}~\bibnamefont
  {Witten}},\ }\href {\doibase 10.1016/0370-2693(81)90885-6} {\bibfield
  {journal} {\bibinfo  {journal} {Phys. Lett. B}\ }\textbf {\bibinfo {volume}
  {105}},\ \bibinfo {pages} {267} (\bibinfo {year}
  {1981}{\natexlab{a}})}\BibitemShut {NoStop}%
\bibitem [{\citenamefont {Witten}(1981{\natexlab{b}})}]{Witten:1981nf}%
  \BibitemOpen
  \bibfield  {author} {\bibinfo {author} {\bibfnamefont {E.}~\bibnamefont
  {Witten}},\ }\href {\doibase 10.1016/0550-3213(81)90006-7} {\bibfield
  {journal} {\bibinfo  {journal} {Nucl. Phys. B}\ }\textbf {\bibinfo {volume}
  {188}},\ \bibinfo {pages} {513} (\bibinfo {year}
  {1981}{\natexlab{b}})}\BibitemShut {NoStop}%
\bibitem [{\citenamefont {Artymowski}\ and\ \citenamefont
  {Ben-Dayan}(2020)}]{Artymowski:2019jlh}%
  \BibitemOpen
  \bibfield  {author} {\bibinfo {author} {\bibfnamefont {M.}~\bibnamefont
  {Artymowski}}\ and\ \bibinfo {author} {\bibfnamefont {I.}~\bibnamefont
  {Ben-Dayan}},\ }\href {\doibase 10.3390/sym12050806} {\bibfield  {journal}
  {\bibinfo  {journal} {Symmetry}\ }\textbf {\bibinfo {volume} {12}},\ \bibinfo
  {pages} {806} (\bibinfo {year} {2020})},\ \Eprint
  {http://arxiv.org/abs/1908.07052} {arXiv:1908.07052 [hep-th]} \BibitemShut
  {NoStop}%
\bibitem [{\citenamefont {Akrami}\ \emph {et~al.}(2020)\citenamefont {Akrami}
  \emph {et~al.}}]{Planck:2018jri}%
  \BibitemOpen
  \bibfield  {author} {\bibinfo {author} {\bibfnamefont {Y.}~\bibnamefont
  {Akrami}} \emph {et~al.} (\bibinfo {collaboration} {Planck}),\ }\href
  {\doibase 10.1051/0004-6361/201833887} {\bibfield  {journal} {\bibinfo
  {journal} {Astron. Astrophys.}\ }\textbf {\bibinfo {volume} {641}},\ \bibinfo
  {pages} {A10} (\bibinfo {year} {2020})},\ \Eprint
  {http://arxiv.org/abs/1807.06211} {arXiv:1807.06211 [astro-ph.CO]}
  \BibitemShut {NoStop}%
\bibitem [{\citenamefont {Lewis}\ \emph {et~al.}(2000)\citenamefont {Lewis},
  \citenamefont {Challinor},\ and\ \citenamefont {Lasenby}}]{Lewis:1999bs}%
  \BibitemOpen
  \bibfield  {author} {\bibinfo {author} {\bibfnamefont {A.}~\bibnamefont
  {Lewis}}, \bibinfo {author} {\bibfnamefont {A.}~\bibnamefont {Challinor}}, \
  and\ \bibinfo {author} {\bibfnamefont {A.}~\bibnamefont {Lasenby}},\ }\href
  {\doibase 10.1086/309179} {\bibfield  {journal} {\bibinfo  {journal}
  {Astrophys. J.}\ }\textbf {\bibinfo {volume} {538}},\ \bibinfo {pages} {473}
  (\bibinfo {year} {2000})},\ \Eprint {http://arxiv.org/abs/astro-ph/9911177}
  {arXiv:astro-ph/9911177} \BibitemShut {NoStop}%
\bibitem [{\citenamefont {Lewis}\ and\ \citenamefont
  {Bridle}(2002)}]{Lewis:2002ah}%
  \BibitemOpen
  \bibfield  {author} {\bibinfo {author} {\bibfnamefont {A.}~\bibnamefont
  {Lewis}}\ and\ \bibinfo {author} {\bibfnamefont {S.}~\bibnamefont {Bridle}},\
  }\href {\doibase 10.1103/PhysRevD.66.103511} {\bibfield  {journal} {\bibinfo
  {journal} {Phys. Rev. D}\ }\textbf {\bibinfo {volume} {66}},\ \bibinfo
  {pages} {103511} (\bibinfo {year} {2002})},\ \Eprint
  {http://arxiv.org/abs/astro-ph/0205436} {arXiv:astro-ph/0205436} \BibitemShut
  {NoStop}%
\bibitem [{\citenamefont {Mortonson}\ \emph {et~al.}(2011)\citenamefont
  {Mortonson}, \citenamefont {Peiris},\ and\ \citenamefont
  {Easther}}]{Mortonson:2010er}%
  \BibitemOpen
  \bibfield  {author} {\bibinfo {author} {\bibfnamefont {M.~J.}\ \bibnamefont
  {Mortonson}}, \bibinfo {author} {\bibfnamefont {H.~V.}\ \bibnamefont
  {Peiris}}, \ and\ \bibinfo {author} {\bibfnamefont {R.}~\bibnamefont
  {Easther}},\ }\href {\doibase 10.1103/PhysRevD.83.043505} {\bibfield
  {journal} {\bibinfo  {journal} {Phys. Rev. D}\ }\textbf {\bibinfo {volume}
  {83}},\ \bibinfo {pages} {043505} (\bibinfo {year} {2011})},\ \Eprint
  {http://arxiv.org/abs/1007.4205} {arXiv:1007.4205 [astro-ph.CO]} \BibitemShut
  {NoStop}%
\bibitem [{\citenamefont {Easther}\ and\ \citenamefont
  {Peiris}(2012)}]{Easther:2011yq}%
  \BibitemOpen
  \bibfield  {author} {\bibinfo {author} {\bibfnamefont {R.}~\bibnamefont
  {Easther}}\ and\ \bibinfo {author} {\bibfnamefont {H.~V.}\ \bibnamefont
  {Peiris}},\ }\href {\doibase 10.1103/PhysRevD.85.103533} {\bibfield
  {journal} {\bibinfo  {journal} {Phys. Rev. D}\ }\textbf {\bibinfo {volume}
  {85}},\ \bibinfo {pages} {103533} (\bibinfo {year} {2012})},\ \Eprint
  {http://arxiv.org/abs/1112.0326} {arXiv:1112.0326 [astro-ph.CO]} \BibitemShut
  {NoStop}%
\bibitem [{\citenamefont {Trotta}(2008)}]{Trotta:2008qt}%
  \BibitemOpen
  \bibfield  {author} {\bibinfo {author} {\bibfnamefont {R.}~\bibnamefont
  {Trotta}},\ }\href {\doibase 10.1080/00107510802066753} {\bibfield  {journal}
  {\bibinfo  {journal} {Contemp. Phys.}\ }\textbf {\bibinfo {volume} {49}},\
  \bibinfo {pages} {71} (\bibinfo {year} {2008})},\ \Eprint
  {http://arxiv.org/abs/0803.4089} {arXiv:0803.4089 [astro-ph]} \BibitemShut
  {NoStop}%
\bibitem [{\citenamefont {Beutler}\ \emph {et~al.}(2011)\citenamefont
  {Beutler}, \citenamefont {Blake}, \citenamefont {Colless}, \citenamefont
  {Jones}, \citenamefont {Staveley-Smith}, \citenamefont {Campbell},
  \citenamefont {Parker}, \citenamefont {Saunders},\ and\ \citenamefont
  {Watson}}]{Beutler:2011hx}%
  \BibitemOpen
  \bibfield  {author} {\bibinfo {author} {\bibfnamefont {F.}~\bibnamefont
  {Beutler}}, \bibinfo {author} {\bibfnamefont {C.}~\bibnamefont {Blake}},
  \bibinfo {author} {\bibfnamefont {M.}~\bibnamefont {Colless}}, \bibinfo
  {author} {\bibfnamefont {D.~H.}\ \bibnamefont {Jones}}, \bibinfo {author}
  {\bibfnamefont {L.}~\bibnamefont {Staveley-Smith}}, \bibinfo {author}
  {\bibfnamefont {L.}~\bibnamefont {Campbell}}, \bibinfo {author}
  {\bibfnamefont {Q.}~\bibnamefont {Parker}}, \bibinfo {author} {\bibfnamefont
  {W.}~\bibnamefont {Saunders}}, \ and\ \bibinfo {author} {\bibfnamefont
  {F.}~\bibnamefont {Watson}},\ }\href {\doibase
  10.1111/j.1365-2966.2011.19250.x} {\bibfield  {journal} {\bibinfo  {journal}
  {Mon. Not. Roy. Astron. Soc.}\ }\textbf {\bibinfo {volume} {416}},\ \bibinfo
  {pages} {3017} (\bibinfo {year} {2011})},\ \Eprint
  {http://arxiv.org/abs/1106.3366} {arXiv:1106.3366 [astro-ph.CO]} \BibitemShut
  {NoStop}%
\bibitem [{\citenamefont {Ross}\ \emph {et~al.}(2015)\citenamefont {Ross},
  \citenamefont {Samushia}, \citenamefont {Howlett}, \citenamefont {Percival},
  \citenamefont {Burden},\ and\ \citenamefont {Manera}}]{Ross:2014qpa}%
  \BibitemOpen
  \bibfield  {author} {\bibinfo {author} {\bibfnamefont {A.~J.}\ \bibnamefont
  {Ross}}, \bibinfo {author} {\bibfnamefont {L.}~\bibnamefont {Samushia}},
  \bibinfo {author} {\bibfnamefont {C.}~\bibnamefont {Howlett}}, \bibinfo
  {author} {\bibfnamefont {W.~J.}\ \bibnamefont {Percival}}, \bibinfo {author}
  {\bibfnamefont {A.}~\bibnamefont {Burden}}, \ and\ \bibinfo {author}
  {\bibfnamefont {M.}~\bibnamefont {Manera}},\ }\href {\doibase
  10.1093/mnras/stv154} {\bibfield  {journal} {\bibinfo  {journal} {Mon. Not.
  Roy. Astron. Soc.}\ }\textbf {\bibinfo {volume} {449}},\ \bibinfo {pages}
  {835} (\bibinfo {year} {2015})},\ \Eprint {http://arxiv.org/abs/1409.3242}
  {arXiv:1409.3242 [astro-ph.CO]} \BibitemShut {NoStop}%
\bibitem [{\citenamefont {Anderson}\ \emph {et~al.}(2014)\citenamefont
  {Anderson} \emph {et~al.}}]{BOSS:2013rlg}%
  \BibitemOpen
  \bibfield  {author} {\bibinfo {author} {\bibfnamefont {L.}~\bibnamefont
  {Anderson}} \emph {et~al.} (\bibinfo {collaboration} {BOSS}),\ }\href
  {\doibase 10.1093/mnras/stu523} {\bibfield  {journal} {\bibinfo  {journal}
  {Mon. Not. Roy. Astron. Soc.}\ }\textbf {\bibinfo {volume} {441}},\ \bibinfo
  {pages} {24} (\bibinfo {year} {2014})},\ \Eprint
  {http://arxiv.org/abs/1312.4877} {arXiv:1312.4877 [astro-ph.CO]} \BibitemShut
  {NoStop}%
\bibitem [{\citenamefont {Liddle}(2007)}]{Liddle:2007fy}%
  \BibitemOpen
  \bibfield  {author} {\bibinfo {author} {\bibfnamefont {A.~R.}\ \bibnamefont
  {Liddle}},\ }\href {\doibase 10.1111/j.1745-3933.2007.00306.x} {\bibfield
  {journal} {\bibinfo  {journal} {Mon. Not. Roy. Astron. Soc.}\ }\textbf
  {\bibinfo {volume} {377}},\ \bibinfo {pages} {L74} (\bibinfo {year}
  {2007})},\ \Eprint {http://arxiv.org/abs/astro-ph/0701113}
  {arXiv:astro-ph/0701113} \BibitemShut {NoStop}%
\bibitem [{\citenamefont {Spiegelhalter}\ \emph {et~al.}(2002)\citenamefont
  {Spiegelhalter}, \citenamefont {Best}, \citenamefont {Carlin},\ and\
  \citenamefont {van~der Linde}}]{Spiegelhalter:2002yvw}%
  \BibitemOpen
  \bibfield  {author} {\bibinfo {author} {\bibfnamefont {D.~J.}\ \bibnamefont
  {Spiegelhalter}}, \bibinfo {author} {\bibfnamefont {N.~G.}\ \bibnamefont
  {Best}}, \bibinfo {author} {\bibfnamefont {B.~P.}\ \bibnamefont {Carlin}}, \
  and\ \bibinfo {author} {\bibfnamefont {A.}~\bibnamefont {van~der Linde}},\
  }\href {\doibase 10.1111/1467-9868.00353} {\bibfield  {journal} {\bibinfo
  {journal} {J. Roy. Statist. Soc. B}\ }\textbf {\bibinfo {volume} {64}},\
  \bibinfo {pages} {583} (\bibinfo {year} {2002})}\BibitemShut {NoStop}%
\bibitem [{\citenamefont {Winkler}\ \emph {et~al.}(2020)\citenamefont
  {Winkler}, \citenamefont {Gerbino},\ and\ \citenamefont
  {Benetti}}]{Winkler:2019hkh}%
  \BibitemOpen
  \bibfield  {author} {\bibinfo {author} {\bibfnamefont {M.~W.}\ \bibnamefont
  {Winkler}}, \bibinfo {author} {\bibfnamefont {M.}~\bibnamefont {Gerbino}}, \
  and\ \bibinfo {author} {\bibfnamefont {M.}~\bibnamefont {Benetti}},\ }\href
  {\doibase 10.1103/PhysRevD.101.083525} {\bibfield  {journal} {\bibinfo
  {journal} {Phys. Rev. D}\ }\textbf {\bibinfo {volume} {101}},\ \bibinfo
  {pages} {083525} (\bibinfo {year} {2020})},\ \Eprint
  {http://arxiv.org/abs/1911.11148} {arXiv:1911.11148 [astro-ph.CO]}
  \BibitemShut {NoStop}%
\bibitem [{\citenamefont {Hazumi}\ \emph {et~al.}(2019)\citenamefont {Hazumi}
  \emph {et~al.}}]{Hazumi:2019lys}%
  \BibitemOpen
  \bibfield  {author} {\bibinfo {author} {\bibfnamefont {M.}~\bibnamefont
  {Hazumi}} \emph {et~al.},\ }\href {\doibase 10.1007/s10909-019-02150-5}
  {\bibfield  {journal} {\bibinfo  {journal} {J. Low Temp. Phys.}\ }\textbf
  {\bibinfo {volume} {194}},\ \bibinfo {pages} {443} (\bibinfo {year}
  {2019})}\BibitemShut {NoStop}%
\bibitem [{\citenamefont {Felder}\ \emph
  {et~al.}(2001{\natexlab{a}})\citenamefont {Felder}, \citenamefont
  {Garcia-Bellido}, \citenamefont {Greene}, \citenamefont {Kofman},
  \citenamefont {Linde},\ and\ \citenamefont {Tkachev}}]{Felder:2000hj}%
  \BibitemOpen
  \bibfield  {author} {\bibinfo {author} {\bibfnamefont {G.~N.}\ \bibnamefont
  {Felder}}, \bibinfo {author} {\bibfnamefont {J.}~\bibnamefont
  {Garcia-Bellido}}, \bibinfo {author} {\bibfnamefont {P.~B.}\ \bibnamefont
  {Greene}}, \bibinfo {author} {\bibfnamefont {L.}~\bibnamefont {Kofman}},
  \bibinfo {author} {\bibfnamefont {A.~D.}\ \bibnamefont {Linde}}, \ and\
  \bibinfo {author} {\bibfnamefont {I.}~\bibnamefont {Tkachev}},\ }\href
  {\doibase 10.1103/PhysRevLett.87.011601} {\bibfield  {journal} {\bibinfo
  {journal} {Phys. Rev. Lett.}\ }\textbf {\bibinfo {volume} {87}},\ \bibinfo
  {pages} {011601} (\bibinfo {year} {2001}{\natexlab{a}})},\ \Eprint
  {http://arxiv.org/abs/hep-ph/0012142} {arXiv:hep-ph/0012142} \BibitemShut
  {NoStop}%
\bibitem [{\citenamefont {Felder}\ \emph
  {et~al.}(2001{\natexlab{b}})\citenamefont {Felder}, \citenamefont {Kofman},\
  and\ \citenamefont {Linde}}]{Felder:2001kt}%
  \BibitemOpen
  \bibfield  {author} {\bibinfo {author} {\bibfnamefont {G.~N.}\ \bibnamefont
  {Felder}}, \bibinfo {author} {\bibfnamefont {L.}~\bibnamefont {Kofman}}, \
  and\ \bibinfo {author} {\bibfnamefont {A.~D.}\ \bibnamefont {Linde}},\ }\href
  {\doibase 10.1103/PhysRevD.64.123517} {\bibfield  {journal} {\bibinfo
  {journal} {Phys. Rev. D}\ }\textbf {\bibinfo {volume} {64}},\ \bibinfo
  {pages} {123517} (\bibinfo {year} {2001}{\natexlab{b}})},\ \Eprint
  {http://arxiv.org/abs/hep-th/0106179} {arXiv:hep-th/0106179} \BibitemShut
  {NoStop}%
\bibitem [{\citenamefont {Garcia-Bellido}\ \emph {et~al.}(2008)\citenamefont
  {Garcia-Bellido}, \citenamefont {Figueroa},\ and\ \citenamefont
  {Sastre}}]{Garcia-Bellido:2007fiu}%
  \BibitemOpen
  \bibfield  {author} {\bibinfo {author} {\bibfnamefont {J.}~\bibnamefont
  {Garcia-Bellido}}, \bibinfo {author} {\bibfnamefont {D.~G.}\ \bibnamefont
  {Figueroa}}, \ and\ \bibinfo {author} {\bibfnamefont {A.}~\bibnamefont
  {Sastre}},\ }\href {\doibase 10.1103/PhysRevD.77.043517} {\bibfield
  {journal} {\bibinfo  {journal} {Phys. Rev. D}\ }\textbf {\bibinfo {volume}
  {77}},\ \bibinfo {pages} {043517} (\bibinfo {year} {2008})},\ \Eprint
  {http://arxiv.org/abs/0707.0839} {arXiv:0707.0839 [hep-ph]} \BibitemShut
  {NoStop}%
\bibitem [{\citenamefont {Antusch}\ \emph {et~al.}(2017)\citenamefont
  {Antusch}, \citenamefont {Cefala},\ and\ \citenamefont
  {Orani}}]{Antusch:2016con}%
  \BibitemOpen
  \bibfield  {author} {\bibinfo {author} {\bibfnamefont {S.}~\bibnamefont
  {Antusch}}, \bibinfo {author} {\bibfnamefont {F.}~\bibnamefont {Cefala}}, \
  and\ \bibinfo {author} {\bibfnamefont {S.}~\bibnamefont {Orani}},\ }\href
  {\doibase 10.1103/PhysRevLett.118.011303} {\bibfield  {journal} {\bibinfo
  {journal} {Phys. Rev. Lett.}\ }\textbf {\bibinfo {volume} {118}},\ \bibinfo
  {pages} {011303} (\bibinfo {year} {2017})},\ \bibinfo {note} {[Erratum:
  Phys.Rev.Lett. 120, 219901 (2018)]},\ \Eprint
  {http://arxiv.org/abs/1607.01314} {arXiv:1607.01314 [astro-ph.CO]}
  \BibitemShut {NoStop}%
\end{thebibliography}%

\end{document}